\documentclass[twocolumn]{aastex631}

\submitjournal{AJ}

\shorttitle{Stellar kinematics in Mon OB1 and R1}
\shortauthors{Lim et al.}
\begin{document}

\title{A Gaia view on the star formation in the Monoceros OB1 and R1 associations}

\correspondingauthor{Beomdu Lim}
\email{blim@kongju.ac.kr}

\author[0000-0001-5797-9828]{Beomdu Lim}
\affiliation{Department of Earth Science Education, Kongju National University, 
56 Gongjudaehak-ro, Gongju-si, Chungcheongnam-do 314701, Republic of Korea}
\affiliation{School of Space Research, Kyung Hee University 1732, 
Deogyeong-daero, Giheung-gu, Yongin-si, Gyeonggi-do 17104, 
Republic of Korea}
\affiliation{Korea Astronomy and Space Science Institute, 776 
Daedeokdae-ro, Yuseong-gu, Daejeon 34055, Republic of Korea}

\author[0000-0003-4071-9346]{Ya\"el Naz\'e}
\thanks{FNRS Senior Research Associate}
\affiliation{Space sciences, Technologies and Astrophysics Research Institute, 
Universit\'e de Li\`ege, Quartier Agora, All\'ee du 6 Ao\^ut 19c, B\^at. B5c, 4000, Li\`ege, Belgium} 

\author[0000-0002-5097-8707]{Jongsuk Hong}
\affiliation{Korea Astronomy and Space Science Institute, 776 Daedeokdae-ro, Yuseong-gu, Daejeon 34055, Republic of Korea}

\author[0000-0001-6216-0462]{Sung-yong Yoon}
\affiliation{School of Space Research, Kyung Hee University 1732, Deogyeong-daero, Giheung-gu, Yongin-si, Gyeonggi-do 17104, Republic of Korea}
\affiliation{Korea Astronomy and Space Science Institute, 776 
Daedeokdae-ro, Yuseong-gu, Daejeon 34055, Republic of Korea}

\author[0000-0003-3651-2924]{Jinhee Lee}
\affiliation{Department of Earth Science, Pusan National University, 2, Busandaehak-ro 63beon-gil, Geumjeong-gu, Busan 46241, Republic of Korea}

\author[0000-0002-2013-1273]{Narae Hwang}
\affiliation{Korea Astronomy and Space Science Institute, 776 Daedeokdae-ro, Yuseong-gu, Daejeon 34055, Republic of Korea}

\author[0000-0002-6982-7722]{Byeong-Gon Park}
\affiliation{Korea Astronomy and Space Science Institute, 776 Daedeokdae-ro, Yuseong-gu, Daejeon 34055, Republic of Korea}

\author[0000-0003-3119-2087]{Jeong-Eun Lee}
\affiliation{School of Space Research, Kyung Hee University 1732, Deogyeong-daero, Giheung-gu, Yongin-si, Gyeonggi-do 17104, Republic of Korea}

%% Note that the \and command from previous versions of AASTeX is now
%% depreciated in this version as it is no longer necessary. AASTeX 
%% automatically takes care of all commas and "and"s between authors names.

%% AASTeX 6.31 has the new \collaboration and \nocollaboration commands to
%% provide the collaboration status of a group of authors. These commands 
%% can be used either before or after the list of corresponding authors. The
%% argument for \collaboration is the collaboration identifier. Authors are
%% encouraged to surround collaboration identifiers with ()s. The 
%% \nocollaboration command takes no argument and exists to indicate that
%% the nearby authors are not part of surrounding collaborations.

%% Mark off the abstract in the ``abstract'' environment. 
\begin{abstract} 
Stellar kinematics provides the key to understanding star formation process. 
In this respect, we present a kinematic study of the Monoceros OB1 (Mon OB1) 
and R1 (Mon R1) associations using the recent Gaia data and radial 
velocities of stars derived from high-resolution spectroscopy and 
the literature. A total of 728 members are selected using the criteria 
based on the intrinsic properties of young stars, parallaxes, and proper motions. 
The spatial distribution and kinematic properties of members show that these 
associations have distinct substructures. In Mon OB1, we find one northern group 
and two southern groups. Mon R1 is composed of three small 
stellar groups that are spatially and kinematically distinct. Some 
stars are found in a halo around these two
associations. We detect patterns of expansion for most stellar groups in the 
associations. In addition, two stellar groups in Mon OB1 show the signature of 
rotation, which provides an important constraint on cluster formation. The star 
formation history of Mon OB1 is slightly revised. Star formation first occurred in 
the southern region and subsequently in the northern region. Recent star-forming 
events ignited deeper into the southern region, while some stars are escaping 
from Mon OB1, forming a halo. Mon R1 might have formed at the same epoch as the formation 
of the northern group in Mon OB1. Given that star formation is taking place 
on different scales along a large arc-like structure, Mon OB1 and Mon R1 
may be the results of hierarchical star formation.
\end{abstract}

%% Keywords should appear after the \end{abstract} command. 
%% The AAS Journals now uses Unified Astronomy Thesaurus concepts:
%% https://astrothesaurus.org
%% You will be asked to selected these concepts during the submission process
%% but this old "keyword" functionality is maintained in case authors want
%% to include these concepts in their preprints.

%% From the front matter, we move on to the body of the paper.
%% Sections are demarcated by \section and \subsection, respectively.
%% Observe the use of the LaTeX \label
%% command after the \subsection to give a symbolic KEY to the
%% subsection for cross-referencing in a \ref command.
%% You can use LaTeX's \ref and \label commands to keep track of
%% cross-references to sections, equations, tables, and figures.
%% That way, if you change the order of any elements, LaTeX will
%% automatically renumber them.
%%
%% We recommend that authors also use the natbib \citep
%% and \citet commands to identify citations.  The citations are
%% tied to the reference list via symbolic KEYs. The KEY corresponds
%% to the KEY in the \bibitem in the reference list below. 

\section{Introduction} \label{sec:1}
Most young stars form in stellar systems such as 
stellar clusters and associations \citep{LL03,PCA03}. It is 
expected that only less than 10\% of young stellar groups 
will remain gravitationally bound clusters \citep{LL03}, so 
a significant portion of field stars may originate from the 
dissolution of such stellar systems \citep{MS78,BPS07}. The 
formation of stellar systems is interconnected in space and 
time, which leads to form larger structures \citep{EEPZ00,G18}. 
Therefore, they are superb laboratories to understand the 
star formation taking place on various spatial scales. 

Stellar associations are, in general, composed of a single 
or multiple stellar clusters (or groups) and a distributed 
young stellar population \citep{B64,KAG08}. In 
addition, the internal structure is tightly 
associated with the kinematics of constituent stars 
\citep{LSB18,LNGR19,LHY20,LNH21}. The morphological 
features and stellar kinematics provide hints in understanding 
the formation process of stellar associations. To obtain such inference, 
the high precision astrometry from the Gaia mission 
\citep{gaia16} is key as it allows us to select genuine members 
spread over a wide region and investigate their kinematics 
in detail, especially when combined with radial velocity (RV) data. 

Monoceros OB1 (Mon OB1) and R1 (Mon R1) are nearby (within 
1 kpc) stellar associations \citep[etc]{vdB66,OMT96,
SBL97,BCM09}. Mon OB1 hosts the active star-forming 
region (SFR) \object{NGC 2264} with numerous 
substructures \citep{SSB09,KFG14}. A number of 
extensive multi-wavelength imaging surveys for this SFR 
have been conducted due to its proximity and low interstellar 
extinction \citep[etc]{SBL97,PSB00,SBC04,FMS06,SBC08,
SSB09,VPS18}. 

Mon OB1 is about 700 -- 800 pc away from 
the Sun \citep{W56,BF71,FvG91,SBL97,KIO14,DMM21}. Its age 
is about 3 Myr with a large spread of 3 -- 5 Myr 
\citep{SB10,VPS18}.

Extensive RV surveys have been performed for young 
stars in \object{NGC 2264}, the core of Mon OB1 \citep{FHS06,THF15}. 
The velocities of young stars follow the velocity field of the remaining 
molecular gas. A group of stars with RVs larger than the systemic 
velocity ($\sim$ 5 km s$^{-1}$) was found toward the O-type 
binary \object{S Monocerotis} \citep{S09}. \citet{THF15} 
claimed that this group might have formed on the far side of 
the remaining gas compressed by the strong wind from the 
massive star. They also reported the presence of another 
group of stars with systematically small RVs. Recently, the 
internal kinematics of this SFR has been investigated with the 
Gaia proper motion (PM) data \citep{KHS19,BKG20}. As a 
result, a pattern of expansion was detected in this stellar 
group at the north of NGC2264.

Mon R1 is about $2^{\circ}$ west away from 
Mon OB1. It is composed of small SFRs surrounded 
by reflection nebulae such as IC 446, IC 447, NGC 2245, 
and NGC 2247. Infrared observations revealed several 
young stellar groups around Herbig Ae/Be stars in Mon R1 
\citep{WL07,GMM09}. Some Herbig-Haro objects 
were also discovered in those groups \citep{MMD21}. The distance 
to Mon R1 was previously determined in the range 
of  660 pc to 715 pc \citep{vdB66,MDM20,DMM21,MMD21}. This 
association has thus been considered to be at the same 
distance as Mon OB1.

The molecular CO line observations of Mon R1 were 
carried out by \citet{KDT79}. They found a semi 
ring-like structure, which is also evident in Planck 
continuum image at 550 \micron \ \citep{BDP20}. 
The overall velocity field is distinguishable from 
that of Mon OB1 (see also \citealt{OMT96}). 
In the direction of Mon R1, \citet{KDT79} 
identified two velocity components at $-1$ 
to 1 km s$^{-1}$ and 3 to 5 km s$^{-1}$. 
The former component is physically associated 
with IC 446 and IC 447, while the latter one belongs 
to NGC 2245 and NGC 2247. Recently, \citet{BDP20} 
considered the former component as a filament and 
discussed the star formation process along the filament in the 
context of the end-dominated collapse model \citep{B83,
PJH11}. However, the kinematic properties 
of young stars in Mon R1 have not yet been studied 
in detail.

In this study, we aim to investigate not only 
the star formation process in Mon OB1 and Mon R1 
but also the physical association between them by 
probing the kinematics of young stars. For this 
purpose, the recent Gaia Early Data Release 3 
(EDR3; \citealt{gedr3}) is used with RV 
data. We describe data and target selection 
in Section~\ref{sec:2}. The scheme of member 
selection is addressed in Section~\ref{sec:3}. In 
Section~\ref{sec:4}, we investigate the substructures 
in the two associations and the kinematic properties 
of young stars in the substructures. Star formation 
history is also inferred from a color-magnitude diagram 
(CMD). Star and cluster 
formation is discussed in Section~{\ref{sec:5}. Our 
results are summarized in Section~\ref{sec:6}. 

\section{Data} \label{sec:2}
\subsection{Selection of member candidates}\label{ssec:21}
A $6^{\circ} \times 6^{\circ}$ region centered at R.A. = 
06$^{\mathrm{h}}$ 36$^{\mathrm{m}}$ $23\fs52$, decl. 
= $+10^{\circ}$ 04$^{\prime}$ $55\farcs7$ (J2000) was 
surveyed. In order to minimize the inclusion of field interlopers 
in the field of view, we first isolated member candidates 
based on the intrinsic properties of young stars as done in 
\citet{LNH21}. 

Early-type (O- and B-type) stars are probable member 
candidates because of their short lifetime, particularly that of 
O-type stars. We compiled lists of early-type stars taken 
from the data bases of MK classifications \citep{WOE00,R03,
MP03,S09,MSM13} and removed some duplicated stars. A total 
of 609 early-type stars were selected as member candidates. 

Low-mass young stellar objects (YSOs) with warm circumstellar 
disks appear bright at infrared wavelengths \citep{L87}. Hydrogen 
recombination lines are observed in emission due to mass 
accretion \citep{MHC98,MHC03,FKvB13}. X-ray is emitted from 
their hot coronal region \citep[etc]{FDM03,CMP12,RN16}. Based on 
these properties, the membership of young stars in NGC 2264 has 
been thoroughly evaluated by a series of multi-wavelength 
studies \citep{SBL97,PSB00,SBC04,SBC08,SSB09}. We built 
a list of 992 member candidates from these studies. 

\begin{figure}[t]
\epsscale{1.0}
\plotone{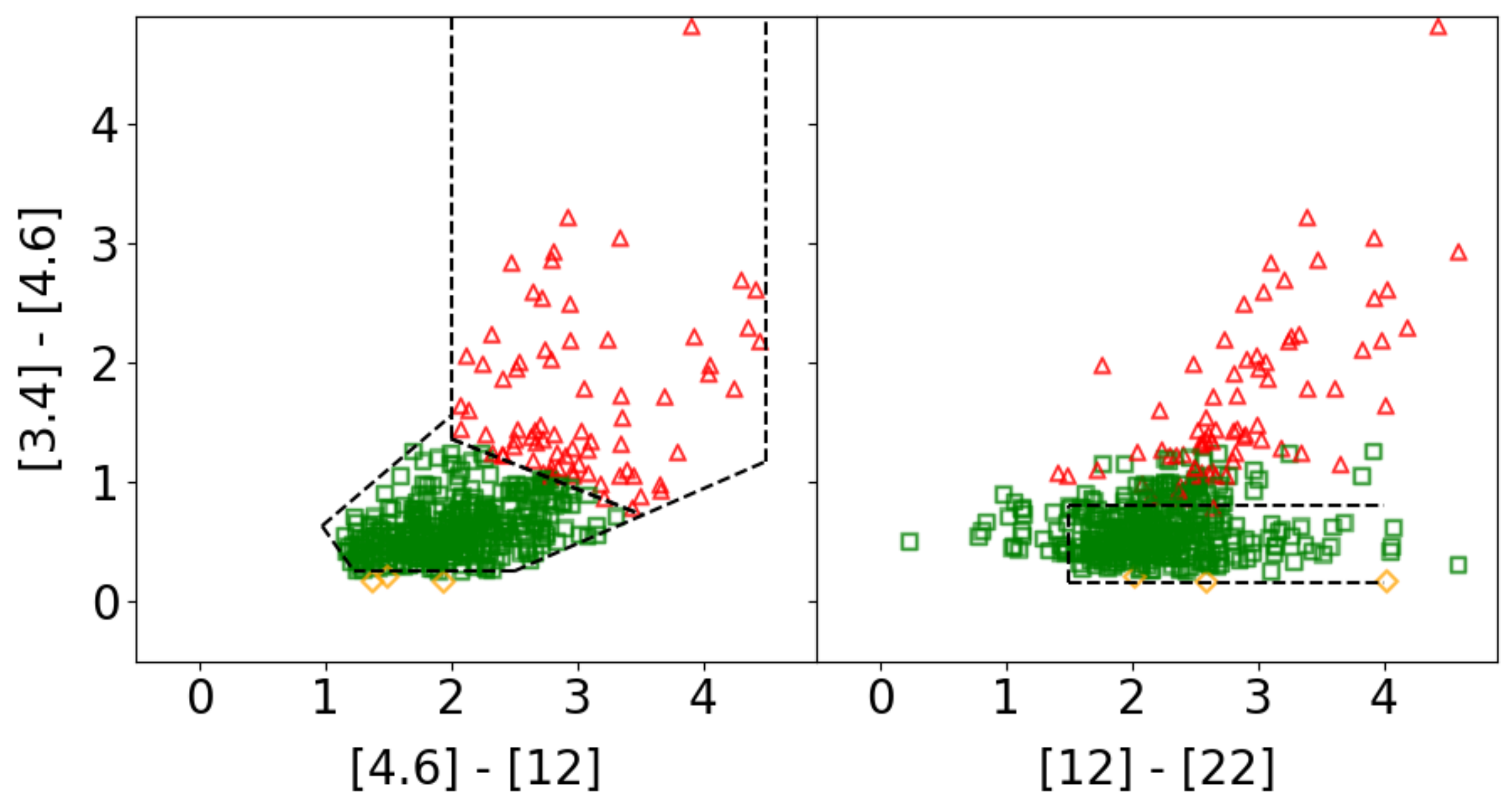}
\caption{Color-color diagrams of YSO candidates classified 
with the AllWISE data \citep{C14}. Red triangle, green square, and 
orange diamond represent Class I, Class II, YSO with a transitional 
disk, respectively. The color criteria for the YSO classification of 
\citet{KL14} are shown by dashed lines. }\label{fig1}
\end{figure}

The Wide-field Infrared Survey Explorer (WISE) has mapped whole 
sky in mid-infrared passbands \citep{WEM10}. We used AllWISE catalogue 
\citep{C14} to identify YSO candidates spread 
over the entire survey region. First, a number of spurious sources were 
rejected by adopting the criterion $ nm/m \leq 0.2$ in each passband 
\citep{KL14}, where $nm$ and $m$ are the number of profile-fit flux 
measurements for a given source with signal-to-noise ratios (SNRs) 
larger than 3 and the total number of profile-fit flux measurements for 
the same source in a given passband, respectively. There could still 
be many spurious sources in W3 and W4 bands. We suppressed those 
sources following \citet{KL14};

\begin{enumerate}
\item W3 band \\
$SNR \geq 5$ \\
$0.45 < \chi^2 < 1.15$  or  $\chi^2 < (SNR - 8)/8 $
\item W4 band\\
$\chi^2 < (2 \times SNR - 20) / 10$
\end{enumerate}

\noindent where $\chi^2$ is the reduced chi-square of profile-fit in 
a given passband. 
  
Additional contaminants such as active galactic nuclei and 
star-forming galaxies were excluded by using the criteria of 
\citet{KL14}. In the end, we classified 75 Class I, 348 Class II, 
and three YSOs with transitional disks according to the scheme 
of \citet{KL14} as shown in Figure~\ref{fig1}. These YSO 
candidates were considered as member candidates. 

We cross-matched the three lists that we built to create a master catalogue 
of member candidates. A total of 29 out of 609 early-type 
stars from the data bases of MK classification were found 
in the member candidate list of NGC 2264 \citep{SBL97,PSB00,SBC04,
SBC08,SSB09}. There are 124 candidates in common between 
the YSO list from the AllWISE data and that of member candidates 
in NGC 2264. Therefore, a total of 302 sources are 
additional YSO candidates throughout our survey region, of which 
two are early-type stars, i.e. they are in the catalogue of 609 
early-type stars. The total number of member candidates is 1872. 

We searched for the counterparts of all member candidates 
in the catalogue of the Gaia EDR3 \citep{gedr3} within a radius 
of $3^{\prime\prime}$. All the OB star candidates were 
found in the Gaia data. The Gaia data further contains 969 
out of 992 member candidates belonging to \object{NGC 2264} 
and 336 out of 426 YSO candidates identified with AllWISE data. 
A total of 1622 candidates brighter than 18 mag in 
$G_{\mathrm{RP}}$ were used for member selection and analysis. 

\subsection{Radial velocity measurements}\label{ssec:22}
\citet{THF15} published the RV data of 695 stars located in 
\object{NGC 2264}. It is noted that the published RVs are actually 
line of sight velocities at the local standard of rest, not 
heliocentric RVs. We took the data and cross-matched them 
with the Gaia catalogue, leading to a total of 684 having 
Gaia counterparts. 

We obtained the high-resolution spectra 
of 14 YSO candidates in Mon R1 using the Immersion GRating Infrared Spectrometer 
(IGRINS, $R \sim 45,000$ -- \citealt{YJB10,PJY14}) attached to 
the 8.2-m Gemini South telescope 
on 2020 February 4, 5, 7, 9, 11, and 12. An ABBA nod technique 
was applied to our observations to subtract the sky background. 
Some A0V stars, such as \object{HIP 30387},  
\object{HIP 36796}, \object{HIP 33297}, and \object{HIP 28686}, 
were observed as telluric standard stars. A large number 
of OH emission lines were observed from blank sky regions for 
wavelength calibration.

Data reduction was performed by using the IGRINS pipeline 
package version 2 \citep{LGK17}\footnote{https://github.com/igrins/plp}. 
This pipeline sequentially executes aperture extraction, the subtraction 
of background emission, bad pixel correction, and wavelength 
calibration. The synthetic spectrum of Vega \citep{CK04} was 
fit to those of the observed A0V stars. The telluric spectra were 
obtained from the spectra of the A0V stars divided by the best-fit 
synthetic spectrum of Vega. Target spectra were corrected by 
the telluric spectra.

Synthetic stellar spectra in the wide temperature range of 3500 to 
9000 K for the solar abundance were generated using {\tt SPECTRUM 
v2.76} \citep{GC94}\footnote{http://www.appstate.edu/~grayro/spectrum/spectrum.html} 
based on a grid of the ODFNEW model atmospheres \citep{CK04}. 
The wavelength of the synthetic spectra in air was converted to 
that in vacuum by using the relation of \citet{C96}. We derived the 
cross-correlation functions between the synthetic spectra and 
the observed spectra of the 14 YSO candidates with {\tt xcsao} task 
in the \textsc{RVSAO} package \citep{KM98}. The velocities at the 
strongest correlation peaks were adopted as RVs. The task 
{\tt xcsao} yields the RV uncertainty based on the $r$ value as 
below \citep{TD79}: 
\begin{center}
\begin{equation}
r = {h \over \sqrt{2}\sigma_a}
\end{equation}
\end{center}

\noindent where $h$ and $\sigma_a$ represent the amplitude of 
a cross-correlation function and the root mean square value of its 
antisymmetric component, respectively. The RV uncertainty is 
then obtained from the relation $3w/8(1+r)$ where $w$ is 
the full width at half maximum of the peak of cross-correlation 
function \citep{KM98}. 

\begin{figure}[t]
\epsscale{1.2}
\plotone{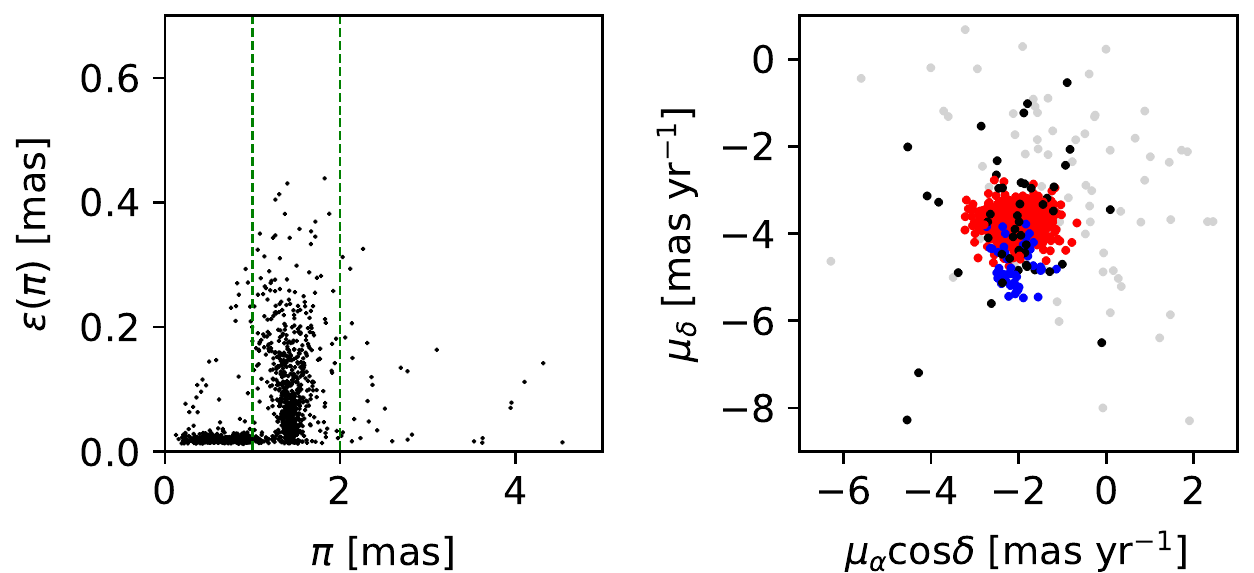}
\caption{Parallax (left) and PM (right) distributions of member candidates. 
The Gaia parallaxes were corrected for zero-point offsets using the recipe 
of \citet{LBB21}. We plot only stars with parallaxes larger than three times 
their associated errors. Dashed lines in the left panel confine the probable 
members to distances between 500 pc and 1000 pc. The right panel 
shows the PM distribution of member candidates between 500 pc and 1000 pc. 
Red, blue, and black dots represent the genuine members of Mon OB1, 
Mon R1, and the halo, respectively, while gray dots denote probable 
nonmembers (see the main text for detail). }\label{fig2}
\end{figure}

Some spectral orders showed poor cross-correlation functions 
because of the small number of lines. The $r$ values tend to be lower 
than 6.0 for these orders. Therefore, we adopted the weighted-mean 
value and standard deviation of RVs measured from spectral 
orders with $r$ larger than 6.0 as the final RV and RV error of a given 
YSO candidate, respectively. The inverse of squared uncertainty was used 
as the weight value. The RVs of YSO candidates were then 
converted to velocities in the local standard of rest frame using 
the \textsc{IRAF}/{\tt rvcorrect} task. 

\subsection{Supplementary data}\label{ssec23}
The Infrared Astronomical Satellite (IRAS) mission surveyed more 
than 95\% of sky at 12, 25, 60, and 100 \micron \ \citep{NHvD84}. Later, 
the Improved Reprocessing of the IRAS Survey (IRIS) provided better 
quality of dust images over the sky \citep{ML05}. In addition, the AKARI 
satellite mapped almost all sky in the four far-infrared bands centered at 
65, 90, 140, and 160\micron \ \citep{DTO15}. 

These infrared maps help to investigate the distribution of interstellar 
material around Mon OB1 and Mon R1. We took the IRIS image at 100$\micron$ 
and the AKARI Far-Infrared Surveyor false-color image of our survey region 
\citep{ML05,DTO15} processed by the {\tt Aladin} interactive sky atlas 
\citep{BFB00,BF14}.

\section{Member selection} \label{sec:3}
We may assume that young stars in an SFR have formed in the same 
molecular cloud. Therefore, they are almost at the same 
distance and share similar kinematic properties. Based on 
this conventional idea, we assessed the membership of the member 
candidates using the Gaia parallax and PM data 
\citep{gedr3}. Systematic zero-point offsets 
that depend on magnitude, color, and position were found in 
the Gaia parallax \citep{LBB21}. We corrected parallaxes for the zero-point 
offsets according to the recipe of \citet[\url{https://gitlab.com/icc-ub/public/gaiadr3_zeropoint}]{LBB21}. 

Stars with parallaxes smaller than three times the associated errors 
were excluded in member selection. In addition, we did not use stars 
with negative parallaxes or close companions (duplication flag = 1), 
or poor astrometric solutions (RUWE $> 1.4$) were not used in analysis 
as well as member selection.
 
The left panel in Figure~\ref{fig2} displays the parallax distributions of 
member candidates. We considered only member candidates between 
500 pc and 1000 pc given the previously determined distances of the 
two associations ($\sim 700$ pc \citealt{vdB66,SBL97,MDM20,MMD21}). 
We plot the PMs of members fulfilling this criterion in the right panel 
of Figure~\ref{fig2}. 

\begin{figure}[t]
\epsscale{1.0}
\plotone{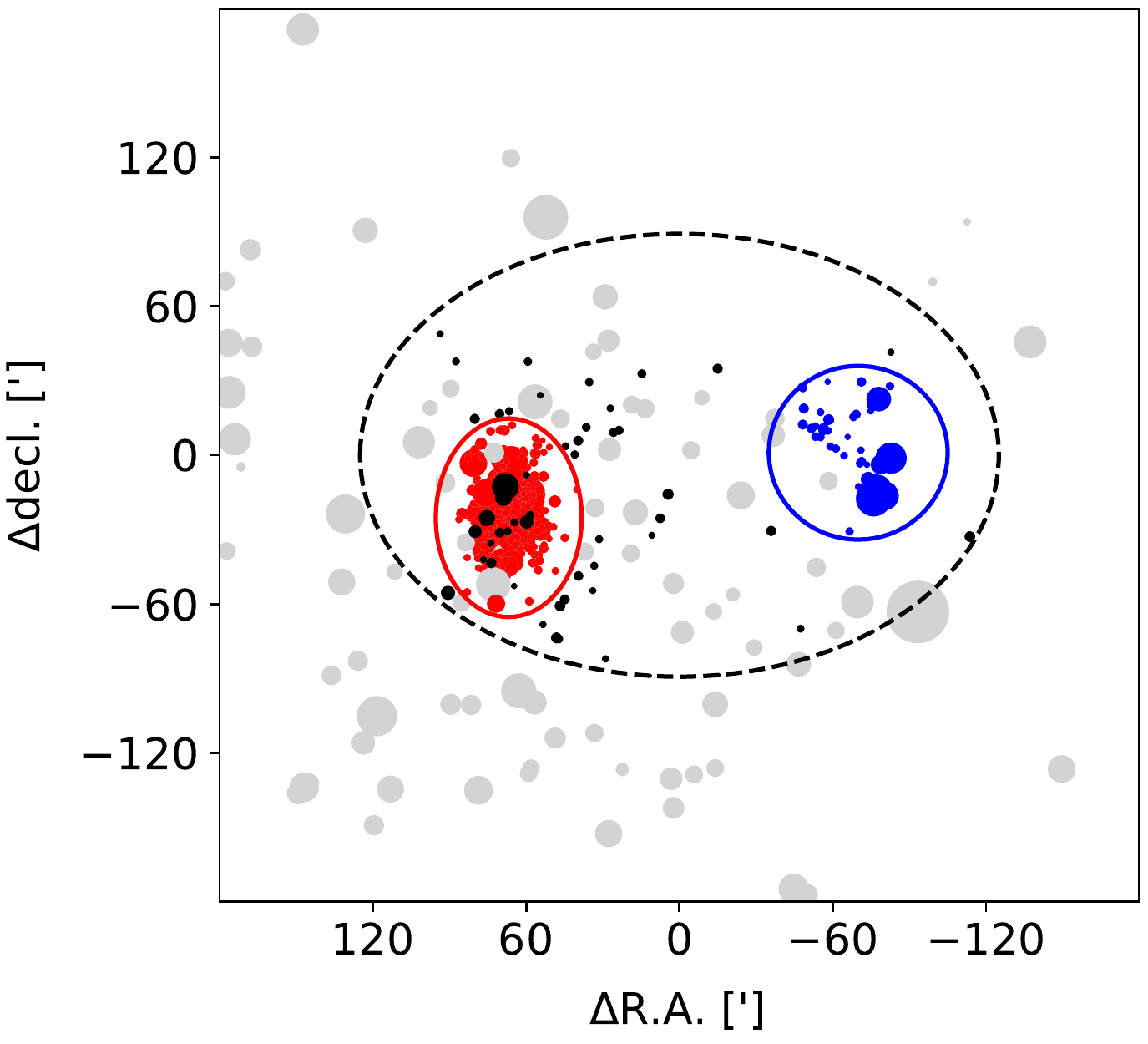}
\caption{Spatial distribution of member candidates. The boundaries of Mon OB1 
and Mon R1 are outlined by a red ellipse and a blue circle, respectively. 
A region encompassing all members is shown by a black ellipse (dashed line). 
The size of dots is proportional to the brightness of stars. The positions 
of stars are relative to the reference coordinate 
R.A. = $06^{\mathrm{h}} \  36^{\mathrm{m}} \ 23\fs52$, 
decl. = $+10^{\circ} \ 04^{\prime}  \ 55\farcs7$ (J2000). The other 
colors of dots are the same as the right panel in Figure~\ref{fig2}.}\label{fig3}
\end{figure}

There are a few groups that have different PMs, on average. These 
groups may be related to Mon OB1 and Mon R1. We assigned the 
member candidates to each association based on the spatial distribution 
in Figure~\ref{fig3}. There are the well-populated association Mon OB1 
to the east and the loose association Mon R1 to the west. We considered 
the boundary of Mon OB1 as a red ellipse with a semi-major axis of 40$^{\prime}$ and 
an eccentricity of 0.7 centered at R.A. = $06^{\mathrm{h}} \  
40^{\mathrm{m}} \ 54\fs56$, decl. = $+09^{\circ} \ 39^{\prime}  \ 43\farcs7$ 
(J2000). The boundary of Mon R1 was assumed to be a blue circle with a 
radius of 35$^{\prime}$ centered at R.A. = $06^{\mathrm{h}} \  
31^{\mathrm{m}} \ 39\fs11$, decl. = $+10^{\circ} \ 05^{\prime}  \ 55\farcs7$ 
(J2000). A total of 653 and 48 stars were found in Mon OB1 and Mon R1, 
respectively, from these criteria.

An iterative process of removing PM outliers was performed to select 
a reliable set of members. We computed the weighted mean values and standard deviations 
of the PMs from the stars in each association. The inverse of squared 
error was adopted as the weight value. Then, stars with PMs within 
the standard deviations ($1\sigma$) from the mean PMs were used to 
determine better mean PMs and their standard deviations. These new 
statistical values were used as the initial values to select 
members. In each region, stars whose PMs are within four times the standard 
deviations ($4\sigma$) from the newly determined mean PMs were selected as 
members. The latter criteria was used to avoid eliminating possible member 
candidates. We redetermined the weighted mean PMs and standard deviations 
using the members. This procedure was repeated until the statistical values 
converge to constant values. The numbers of the final members in Mon OB1 
and Mon R1 are 631 and 46 in total, and they are shown by 
red and blue dots in Figures~\ref{fig2} and \ref{fig3}, respectively.

\begin{figure}[t]
\epsscale{1.0}
\plotone{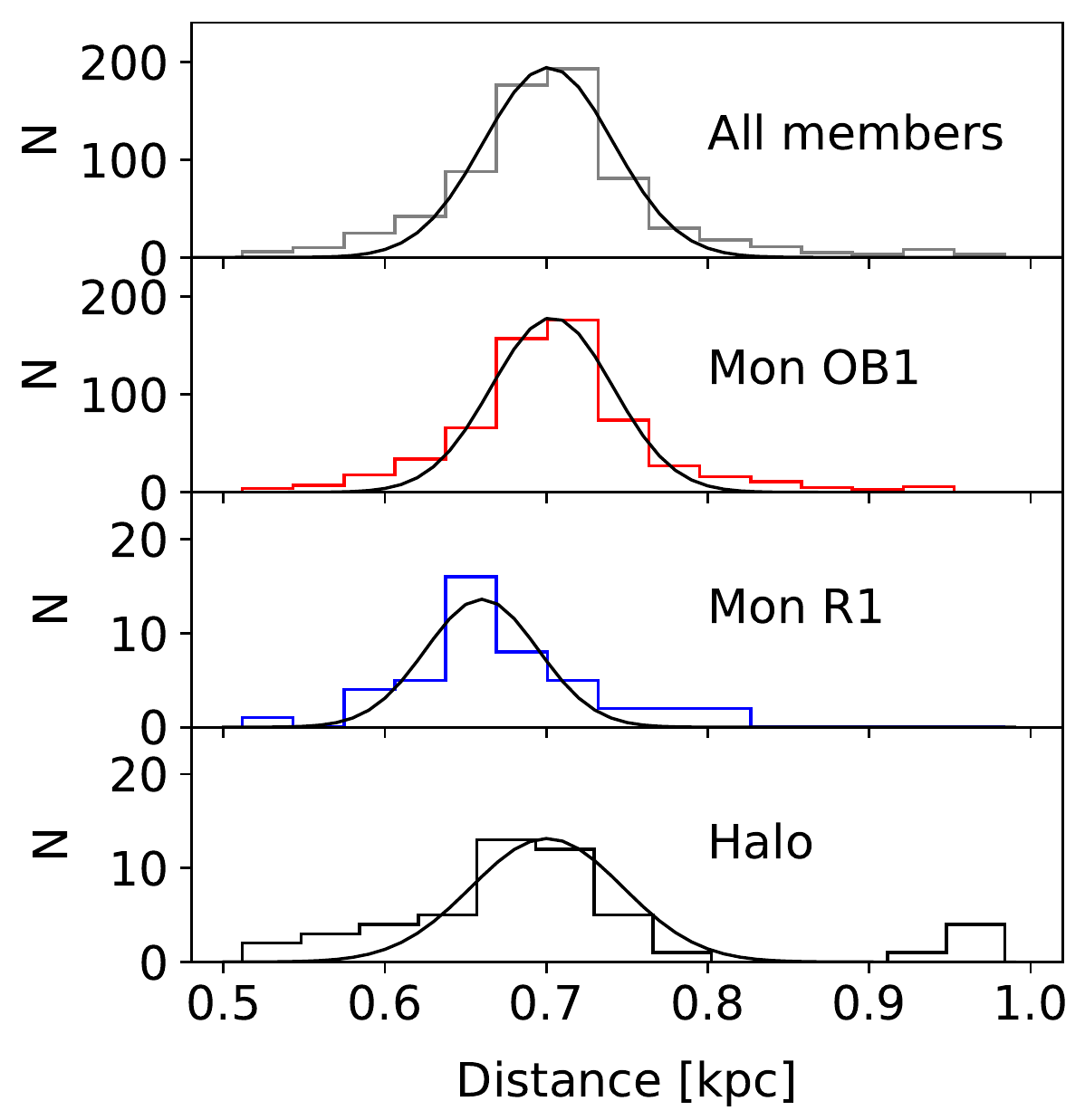}
\caption{Distance distributions. The panels from top to bottom display 
the distance distributions of members in the entire survey region, Mon OB1, 
Mon R1, and the halo, respectively. Distances were obtained from the inverse of 
the Gaia parallaxes \citep{gedr3} after correction for the zero-point offsets 
\citep{LBB21}. We used stars with parallaxes larger than five times their 
associated errors. The black curves shows the best-fit Gaussian distributions. }\label{fig4}
\end{figure}

A total of 53 YSO candidates were also found between the two 
associations. All of them are sparsely distributed within an elliptical 
region (dashed curve in Figure~\ref{fig3}). We refer to this low-stellar 
density region as halo and assume that there is no member outside 
the halo. The YSO members were selected in the same manner as 
above. As a result, we selected 51 YSO members in the halo. The 
halo YSO members are shown by black dots in Figures~\ref{fig2} and ~\ref{fig3}. 

There are a number of early-type stars that do not belong to the two 
associations. These stars, on average, have larger PMs and PM 
dispersion than those of genuine members within the associations 
(most gray dots in the right panel of Figure~\ref{fig2}). In addition, they 
are uniformly distributed over the surveyed region, except for the western 
part obscured by dark clouds. These facts implies that most of them 
may not be genuine members. Therefore, we did not additionally select 
early-type members in the halo.

A total of 728 stars (631 in Mon OB1, 46 in Mon R1, and 51 
in the halo) were finally selected as genuine association members. 
The list of members is presented in Table~\ref{tab1}. We plot 
the distance distributions of these stars in Figure~\ref{fig4}. The distance 
distribution of each group was fit by a Gaussian distribution. The distances 
to Mon OB1, Mon R1, and the halo were determined to be $704\pm38$ 
(s.d.) pc, $660\pm35$ (s.d.) pc, and $700\pm47$ (s.d.) pc, respectively, 
from the best-fit Gaussian distributions. These results are consistent 
with those of previous studies \citep{BF71,FvG91,SBL97,vdB66,MDM20,
DMM21,MMD21} within errors. The members of Mon R1 are systematically 
closer than Mon OB1 although there is only an 1$\sigma$
level difference.

\begin{deluxetable}{lccccccccccccccccc}
\rotate
\setlength\tabcolsep{1.5pt}
\tabletypesize{\tiny}
\tablewidth{0pt}
\tablecaption{List of members \label{tab1}}
\tablehead{\colhead{Sq.} & \colhead{R.A. (2000)} & \colhead{decl. (2000)} & \colhead{$\pi$} & \colhead{$\epsilon(\pi)$} &
\colhead{$\mu_{\alpha}\cos\delta$}  & \colhead{$\epsilon(\mu_{\alpha}\cos\delta)$} & \colhead{$\mu_{\delta}$} & \colhead{$\epsilon(\mu_{\delta})$} & \colhead{$G$}  &
\colhead{$G_{BP}$}  & \colhead{$G_{RP}$}  & \colhead{$G_{BP}-G_{RP}$} & \colhead{RV} & \colhead{$\epsilon$(RV)} & \colhead{Member type} & \colhead{Region} & \colhead{Group} \\
 & \colhead{(h:m:s)} & \colhead{($\degr:\arcmin:\arcsec$)} &  \colhead{(mas)} & \colhead{(mas)} & \colhead{(mas yr$^{-1}$)} &\colhead{(mas yr$^{-1}$)} &
\colhead{(mas yr$^{-1}$)} &\colhead{(mas yr$^{-1}$)} & \colhead{(mag)} &  \colhead{(mag)} & \colhead{(mag)} &
\colhead{(mag)} & \colhead{(km s$^{-1}$)} & \colhead{(km s$^{-1}$)} & & & }
\startdata
   1 & 06:28:42.43 & +09:32:09.3 & 1.5218 & 0.0304 & -2.623 & 0.034 & -5.604 & 0.030 & 15.4340 & 16.3833 & 14.3820 &  2.0014 & \nodata &\nodata& Y & Halo    &          \\
   2 & 06:30:47.07 & +09:51:54.4 & 1.6008 & 0.0665 & -2.078 & 0.067 & -5.383 & 0.057 & 16.7430 & 18.0777 & 15.5531 &  2.5247 &    1.53 &  0.86 & Y & Mon R1  & IC 447   \\
   3 & 06:30:47.06 & +10:03:46.4 & 1.6570 & 0.0436 & -2.283 & 0.044 & -5.091 & 0.035 &  9.4031 &  9.2126 &  9.2965 & -0.0840 & \nodata &\nodata& E & Mon R1  & IC 447   \\
   4 & 06:30:46.56 & +10:46:28.6 & 1.5125 & 0.1434 & -1.289 & 0.178 & -4.872 & 0.132 & 18.0089 & 19.9074 & 16.6634 &  3.2440 & \nodata &\nodata& Y & Halo    &          \\
   5 & 06:30:48.21 & +09:46:03.7 & 1.7047 & 0.0551 & -2.227 & 0.054 & -5.440 & 0.047 & 16.3639 & 17.4478 & 15.2790 &  2.1687 & \nodata &\nodata& Y & Mon R1  & IC 447   \\
   6 & 06:30:48.19 & +10:32:50.1 & 1.5169 & 0.0774 & -1.495 & 0.088 & -4.847 & 0.075 & 17.0090 & 18.4251 & 15.7492 &  2.6759 & \nodata &\nodata& Y & Mon R1  & IC 446   \\
   7 & 06:30:57.92 & +09:48:34.6 & 1.5884 & 0.0251 & -2.102 & 0.024 & -5.125 & 0.021 &  9.7876 &  9.8408 &  9.6407 &  0.2001 & \nodata &\nodata& E & Mon R1  & IC 447   \\
   8 & 06:31:02.58 & +09:59:20.6 & 1.5790 & 0.1324 & -2.268 & 0.129 & -4.552 & 0.106 & 17.8270 & 18.8366 & 16.5279 &  2.3086 & \nodata &\nodata& Y & Mon R1  & IC 447   \\
   9 & 06:31:03.63 & +10:01:13.6 & 1.5373 & 0.0168 & -2.363 & 0.016 & -4.934 & 0.013 & 11.9248 & 12.3280 & 11.3442 &  0.9838 & \nodata &\nodata& Y & Mon R1  & IC 447   \\
  10 & 06:31:06.12 & +10:27:34.1 & 1.5698 & 0.0160 & -2.647 & 0.019 & -4.337 & 0.016 & 10.8878 & 11.2365 & 10.3401 &  0.8963 & \nodata &\nodata& E & Mon R1  & IC 446   \\ 
 \enddata
\tablecomments{Column (1) : Sequential number. Columns (2) and (3) : The equatorial coordinates of members. Columns (4) and (5) : Absolute parallax and its 
standard error. Columns (6) and (7) : PM in the direction of right ascension and its standard error. Columns (8) and (9): PM in the direction of declination 
and its standard error. Columns (10) -- (12) : Magnitudes in $G$, $G_{BP}$, and $G_{RP}$ bands. Column (13) : $G_{BP} - G_{RP}$ color index. Columns (14) and (15) : Radial 
velocity at the local standard of rest and its error. Column (16) : Member type. `E' represents O- or B-type stars obtained from the data bases of MK classification 
\citep{WOE00,R03,MP03,S09,MSM13}. 'Y' denotes young stellar objects or pre-main sequence members. Column (17) : Region names. Column (18) : Group names. 
The parallax and PM were taken from the Gaia Early Data Release 3 \citep{gedr3}. We corrected for the zero-point offsets for the Gaia parallaxes according to the 
recipe of \citet{LBB21}.  The radial velocities were obtained from \citet{THF15} and our observation. The full table is available electronically.}
\end{deluxetable}

\section{Results}\label{sec:4}
\subsection{Substructures}\label{ssec:41}
A number of previous studies have probed substructures in many 
SFRs from the spatial distributions of stars. However, in the 
absence of kinematic information, it is unclear whether substructures 
are real physical systems. In this study, we search for substructures 
using the Gaia PM and RV data, as well as the spatial distribution of 
members.

\subsubsection{Mon OB1}\label{sssec:411}
A deep photometric study of \citet{SBC08} in optical passbands 
showed that \object{NGC 2264} is composed of two active SFRs 
and a halo. The northern group is located around the massive 
O-type binary \object{S Monocerotis}, while 
the southern group is in the vicinity of the \object{Cone 
Nebula}. A halo surrounds these two SFRs. Later, Spitzer 
observations revealed that the southern group is composed 
of two subgroups of YSOs (Spokes and Cone(C), \citealt{TLY06,SSB09}). 
Additional smaller-scale substructures were identified in that 
southern group \citep{KFG14}.

We searched for substructures in Mon OB1 from the correlations 
between PMs and positions of members. As a result, three stellar 
groups that are spatially and kinematically distinct were identified 
from the correlation between $\mu_{\alpha}\cos\delta$ (PM along 
R.A.) and declination in the upper left panel of Figure~\ref{fig5}. 
We checked the spatial distributions of members within the three 
ellipses in the figure and found that members were populated in 
three specific regions, i.e. they do not show a random distribution 
across this association. We divided members into three groups 
by three solid lines as shown in Figure~\ref{fig5}. These lines 
were adjusted through the visual inspection of their spatial distributions. 
Note that this division could be somewhat arbitrary 
because it is difficult to define the exact boundaries of each group. 

\begin{figure}[t]
\epsscale{1.0}
\plotone{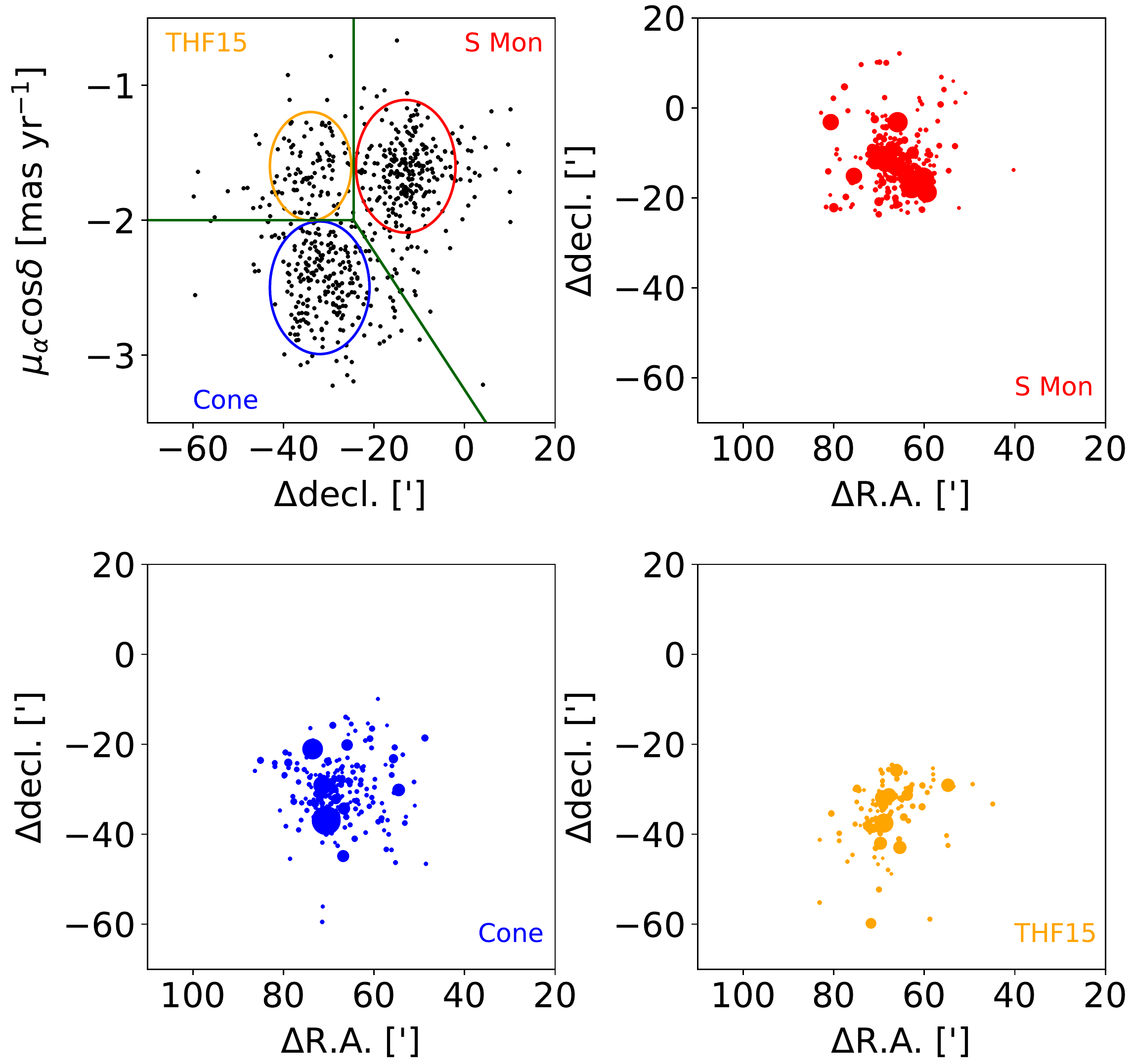}
\caption{Kinematic substructures in Mon OB1. The upper 
left panel shows the correlation between PMs along R. A. and 
positions along declination. It is evident that there are three groups 
of stars that are spatially and kinematically distinct. Three ellipses 
were used to ascertain the spatial distributions of members within them. 
The solid lines represent the arbitrary boundaries of each group. 
The other panels display the spatial distributions of members belonging 
to each group. Red, blue, and orange dots represent the members 
of the S Mon group, the Cone group, and the THF15 group, respectively. 
The size of dots is proportional to the brightness of individual stars. }\label{fig5}
\end{figure}

\begin{figure}[t]
\epsscale{1.0}
\plotone{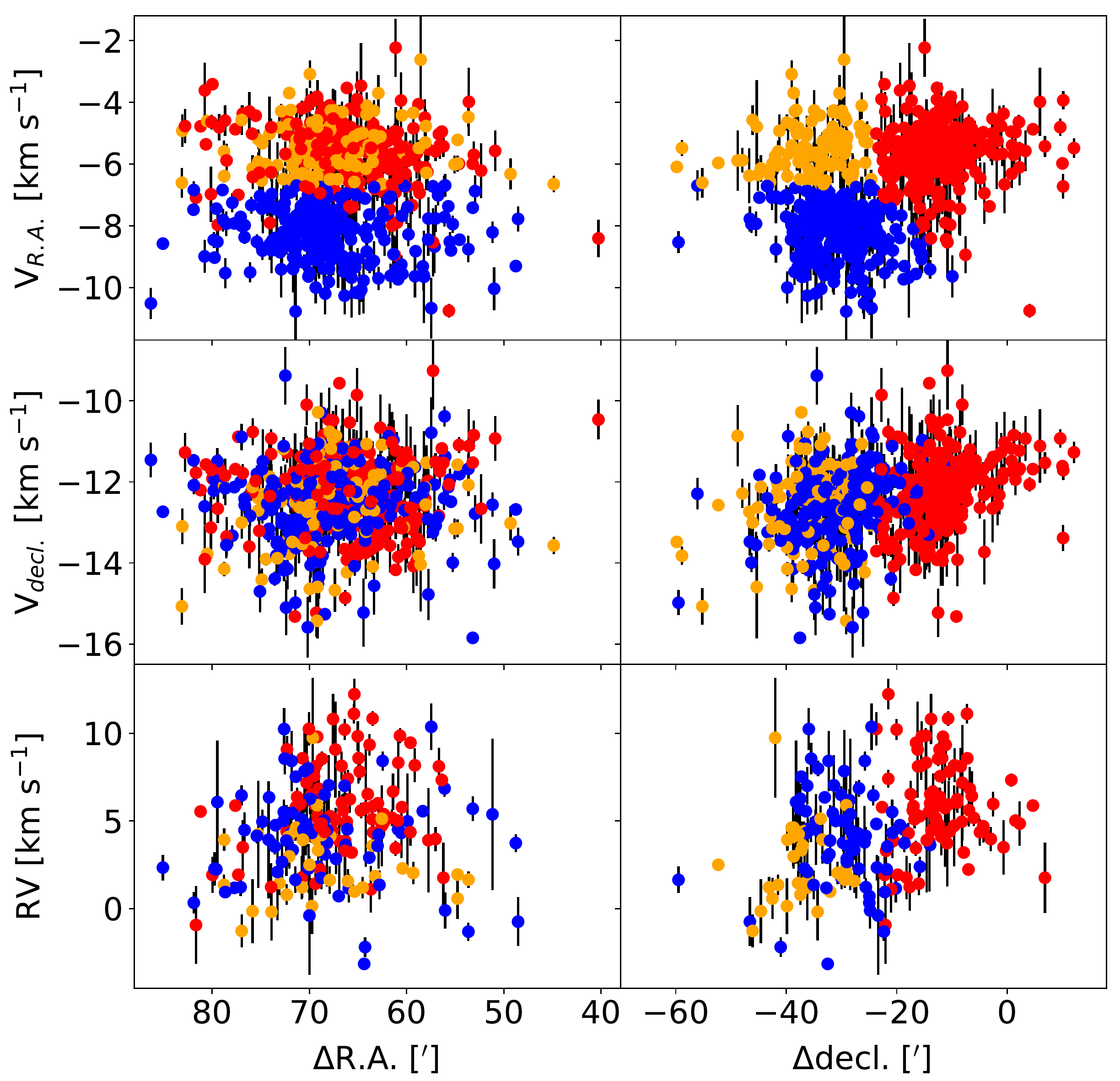}
\caption{Tangential velocities and RVs of stars with respect to R.A. 
and declination in Mon OB1. The vertical lines represent the errors 
of velocities. The colors of symbols are the same as 
those in Figure~\ref{fig5}.
 }\label{fig6}
\end{figure}

The northern group (hereafter S Mon group, red symbol in 
Figure~\ref{fig5}) has a median PM ($\mu_{\alpha}\cos\delta$, 
$\mu_{\delta}$) of ($-1.649$ mas yr$^{-1}$, $-3.655$ mas yr$^{-1}$). 
We could not find additional clustering in the group. On 
the other hand, the southern group is separated into two groups. 
The group shown by blue symbol (the lower-left panel) has a 
median PM of ($-2.455$ mas yr$^{-1}$, $-3.721$ mas yr$^{-1}$): 
we refer to this southern group as the Cone group according 
to the nomenclature of \citet{SBC08}. The other group 
(orange symbol in the lower-right panel) has a median 
PM of ($-1.676$ mas yr$^{-1}$, $-3.685$ mas yr$^{-1}$). This 
group seems to correspond to the blueshifted population 
reported by \citet{THF15}, and therefore we refer to this 
group as THF15 from the names of the authors. There is no 
smaller-scale substructure within 
these two groups as well given the absence of additional 
enhancement in stellar surface density. The central coordinates of these three 
groups were obtained from the median positions of the relevant 
members. We summarize the basic properties of 
these groups in Table~\ref{tab2}. 

Figure~\ref{fig6} displays the tangential velocities ($V_{\mathrm{R. A.}}$ 
and $V_{\mathrm{decl.}}$) and RVs of members. The tangential 
velocities were obtained from the PMs multiplied by the distance 
of 704 pc. Stars in the S Mon and THF15 groups have similar 
$V_{\mathrm{R. A.}}$ but it differs from that of the Cone group, 
while stars in the groups have almost same $V_{\mathrm{decl.}}$ 
(see also Table~\ref{tab2}). The kinematic substructures can also 
be confirmed from the correlations between RVs and positions of 
members. Especially, there is a gradient of RVs with respect to declination ($\sim$ 
0.4 km s$^{-1}$ pc$^{-1}$). The median RVs of the S Mon group, 
Cone, and THF15 are about 5.7, 4.3, and 2.3 km s$^{-1}$, 
respectively. The members in THF15 have RVs systematically 
smaller than the others as reported by \citet{THF15}.

\begin{deluxetable*}{lcccccccc}
\tabletypesize{\tiny}
\tablewidth{0pt}
\tablecaption{Basic properties of the identified stellar groups. \label{tab2}}
\tablehead{\colhead{Group} & \colhead{R.A. (2000)} & \colhead{decl. (2000)} & \colhead{$\mu_{\alpha}\cos\delta$} & \colhead{$\mu_{\delta}$} & \colhead{$V_{\mathrm{R.A.}}$}  & \colhead{$V_{\mathrm{decl.}}$} &
\colhead{RV} & \colhead{$N_{\mathrm{star}}$} \\
 & \colhead{(h:m:s)} & \colhead{($\degr:\arcmin:\arcsec$)} &  \colhead{(mas yr$^{-1}$)} & \colhead{(mas $^{-1}$)} & \colhead{(km s$^{-1}$)} &\colhead{(km s$^{-1}$)} & \colhead{(km s$^{-1}$)} &  }
\startdata
S Mon & 06:40:52.99 & +09:52:25.0 &  -1.649 &  -3.655 & -5.5 & -12.2 & 5.7 & 279\\
Cone & 06:41:02.78 & +09:34:09.9 & -2.455 & -3.721 & -8.2 & -12.4 & 4.3 & 240\\
THF15 & 06:41:03.02 & +09:30:23.0 & -1.676 & -3.685 &-5.6 &-12.3& 2.3 & 112 \\
IC 447 &06:31:14.03 & +09:53:41.8 & -2.143 & -5.137 &-6.7 &-16.1& 1.6 & 18\\
N2245/47 & 06:32:34.72 & +10:15:41.2 &-1.830 & -4.433 &-5.7 & -13.9& 4.5 & 20\\
IC 446 & 06:31:19.60 & +10:24:42.1 & -2.322 & -4.514 & -7.3 & -14.1 & \nodata & 8\\ 
\enddata
\tablecomments{All the shown measurements correspond to median values.}
\end{deluxetable*}

\begin{figure}[t]
\epsscale{1.0}
\plotone{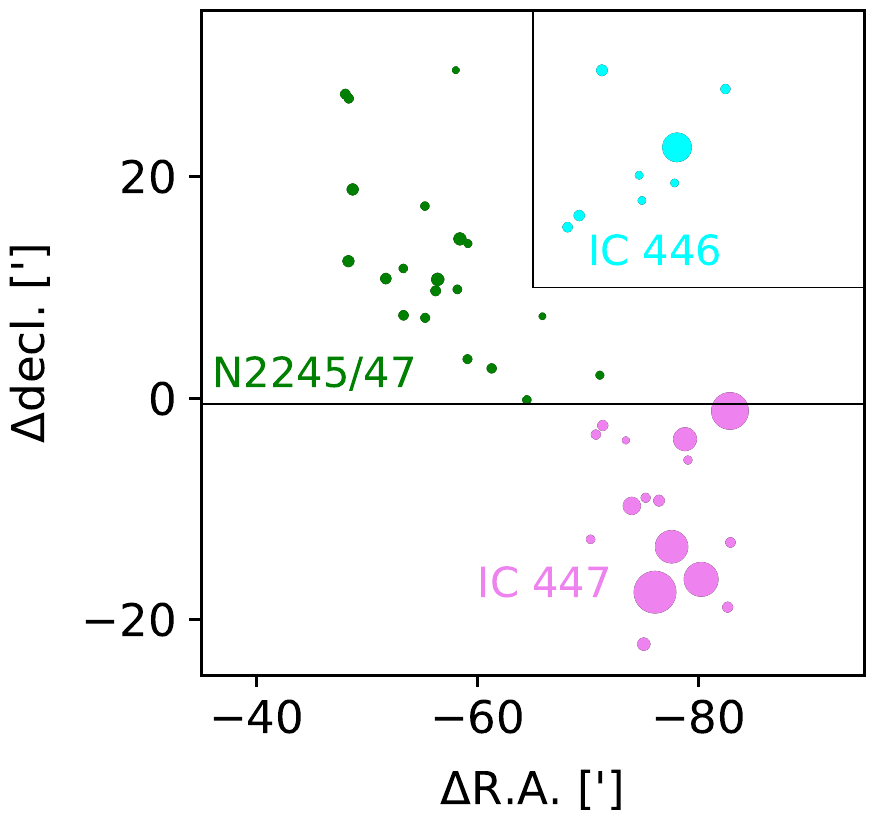}
\caption{Spatial distribution of members in Mon R1. Pink, 
green, and cyan dots represent the members of IC 447, 
N2245/47, and IC 446, respectively. The size of dots is proportional 
to the brightness of individual stars. Solid lines are used to separate 
stellar groups.}\label{fig7}
\end{figure}

\begin{figure}[t]
\epsscale{1.0}
\plotone{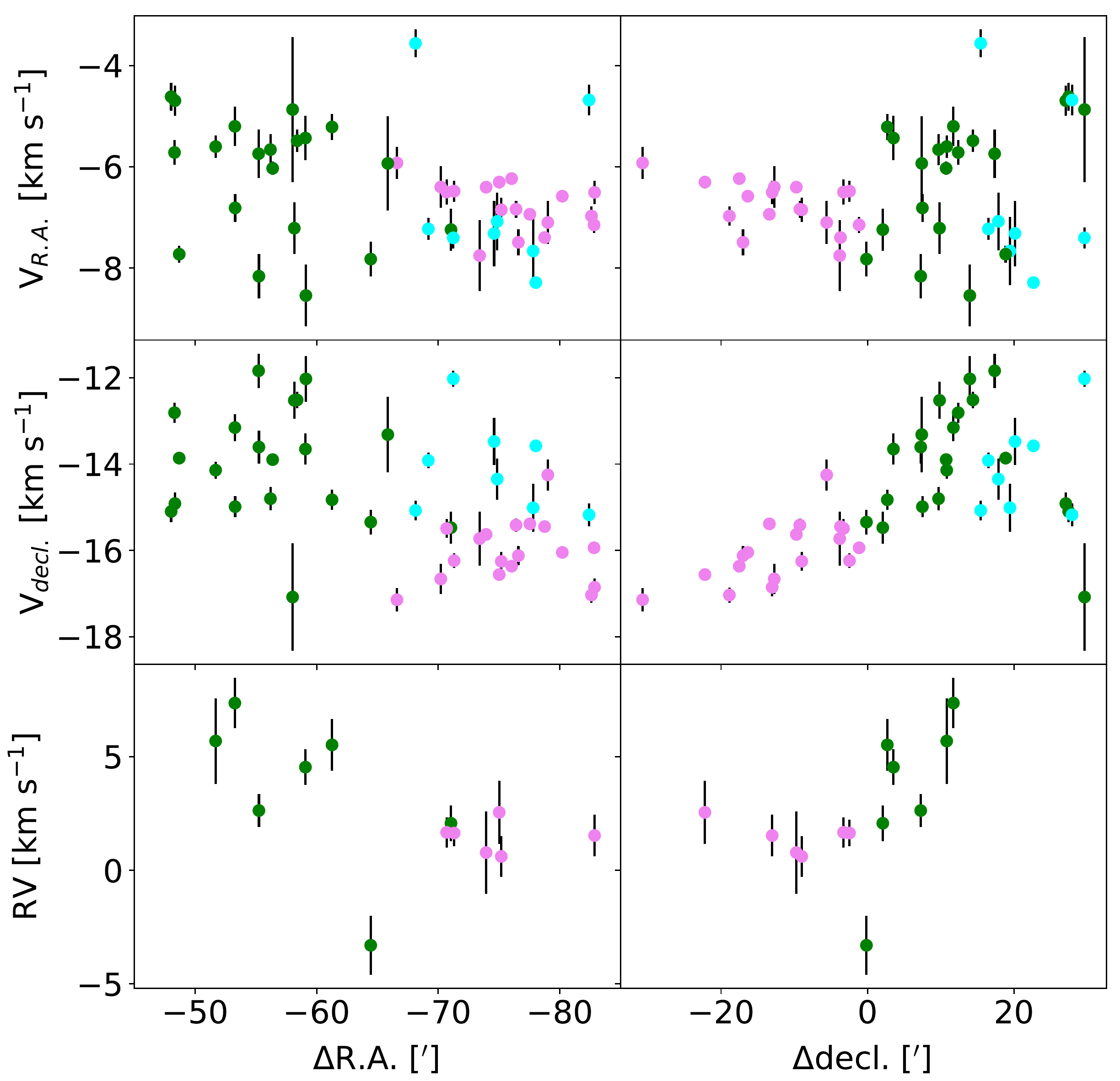}
\caption{Tangential velocities and RVs of stars with respect to R.A. 
and declination in Mon R1. The vertical lines represent the errors 
of velocities.The colors of symbols are the same as 
those in Figure~\ref{fig7}. }\label{fig8}
\end{figure}

\subsubsection{Mon R1}\label{sssec:412}
Figure~\ref{fig7} displays the spatial distribution 
of stars in Mon R1. This stellar association is composed of three small 
stellar groups associated with reflection nebulae. The young open 
cluster Collinder 95 occupies the southern part of Mon R1. The 
reflection nebula IC 447 may have been formed by the early-type 
members of this cluster. We referred to this group as IC 447. 
A partially embedded group of stars in the 
vicinity of the reflection nebula IC 446 are found to the northwest of this 
association (hereafter IC 446). The other embedded group lies between 
the reflection nebulae \object{NGC 2245} and \object{NGC 2247} in the 
eastern region (hereafter N2245/47). 

The boundaries of these groups were determined from the spatial 
distribution of members. Members below the declination of 
$\Delta\mathrm{decl.} = -0\farcm5$ were assigned to the members 
of IC 447. The members of IC 446 were confined to the northwestern 
stars ($\Delta\mathrm{R.A.} < -65\farcm0$ and $\Delta\mathrm{decl.} 
> 10\farcm0$). The rest of stars were considered as the members 
of NGC 2245/47. 

 \begin{figure*}[t]
\epsscale{1.0}
\plotone{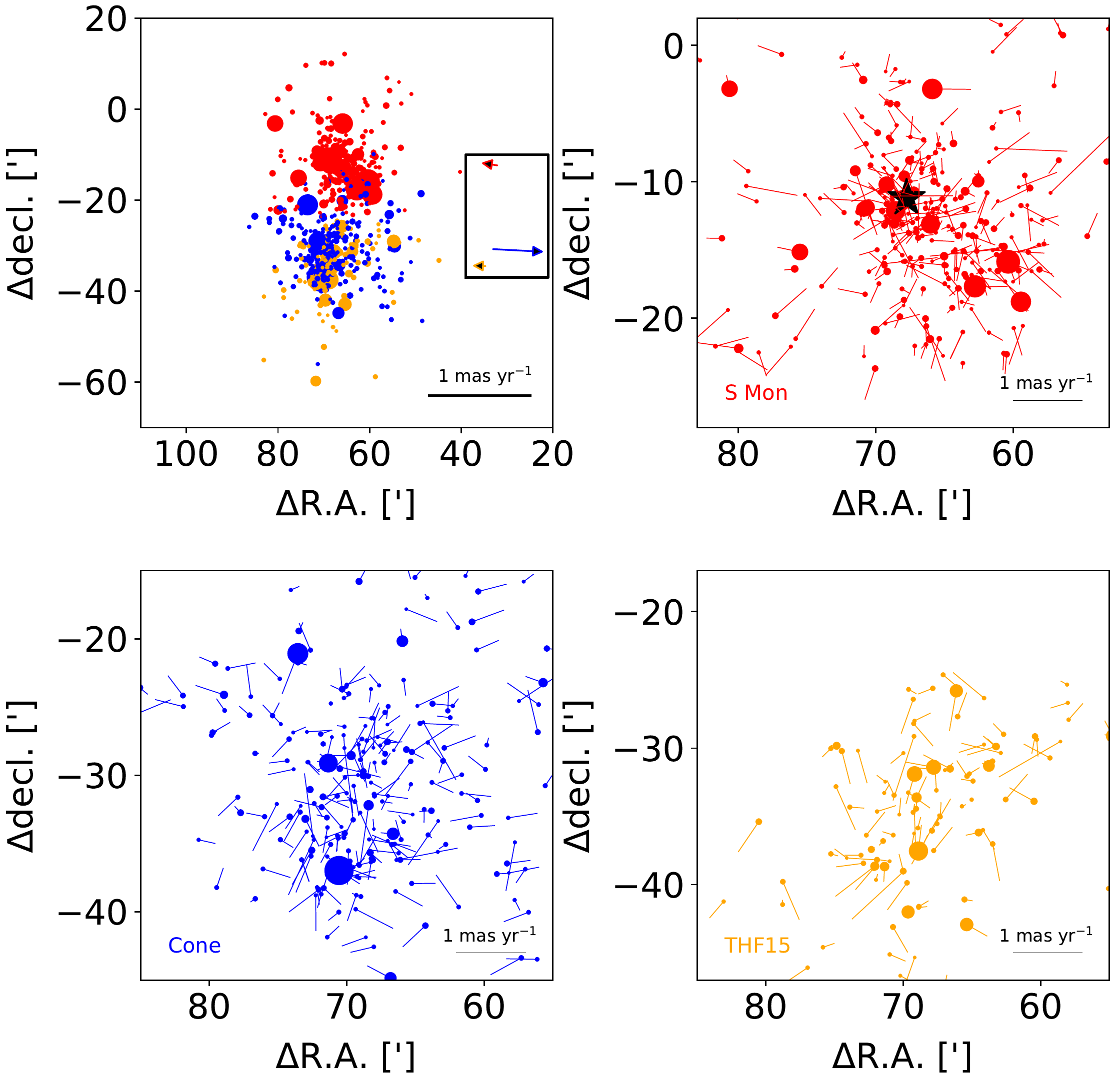}
\caption{Relative PMs of stars in Mon OB1. The upper left panel 
shows the spatial distribution of its three stellar groups. Red, blue, and 
orange dots represent the S Mon group, the Cone group, and 
the THF15 group, respectively. The systemic motions of these groups 
relative to the median PM of Mon OB1 are shown by arrows with the 
same colors in the box, where the PM vectors were shifted along 
R.A. by 35$^{\prime}$ to avoid confusion. We plot the PM vectors (solid lines) of 
individual members relative to the systemic motion of a given group 
in the other panels. The star symbol denotes the O-type binary S Monocerotis. 
The size of dots is scaled by the brightness of individual stars. }\label{fig9}
\end{figure*}

We plot $V_{\mathrm{R. A.}}$, $V_{\mathrm{decl.}}$, and RVs of the 
Mon R1 members in Figure~\ref{fig8}. The three stellar groups 
show somewhat complicated kinematic substructures. The members 
of N2245/47 and IC 446 show a large scatter in tangential velocities 
compared to the members of IC 447. Also, the $V_{\mathrm{decl.}}$ 
of stars continuously vary with declination from IC 447 to IC 446 
($\sim$ 0.5 km s$^{-1}$ pc$^{-1}$). The YSOs with 
RV measurements are found in only the two groups IC 447 and 
N2245/47, and their total number is somewhat limited to probe 
the global variation. Nevertheless, the RVs of the members 
show a tendency similar to the tangential velocities. The 
RVs of IC 447 and N2245/47 follow the velocity fields of 
the remaining molecular gas. The former and latter correspond 
to the gas components in the RV ranges of $-4$ to 2 km s$^{-1}$ and 
2 to 10 km s$^{-1}$ \citep{BDP20}, respectively. It suggests that 
the complicated kinematics of stars seen in the tangential velocity 
distributions might have been inherited from that of their natal cloud. 
We determined the central coordinates of the three groups from the median 
positions of the associated members. Their basic properties 
are summarized in Table~\ref{tab2}.
 
 \begin{figure}[t]
\epsscale{1.0}
\plotone{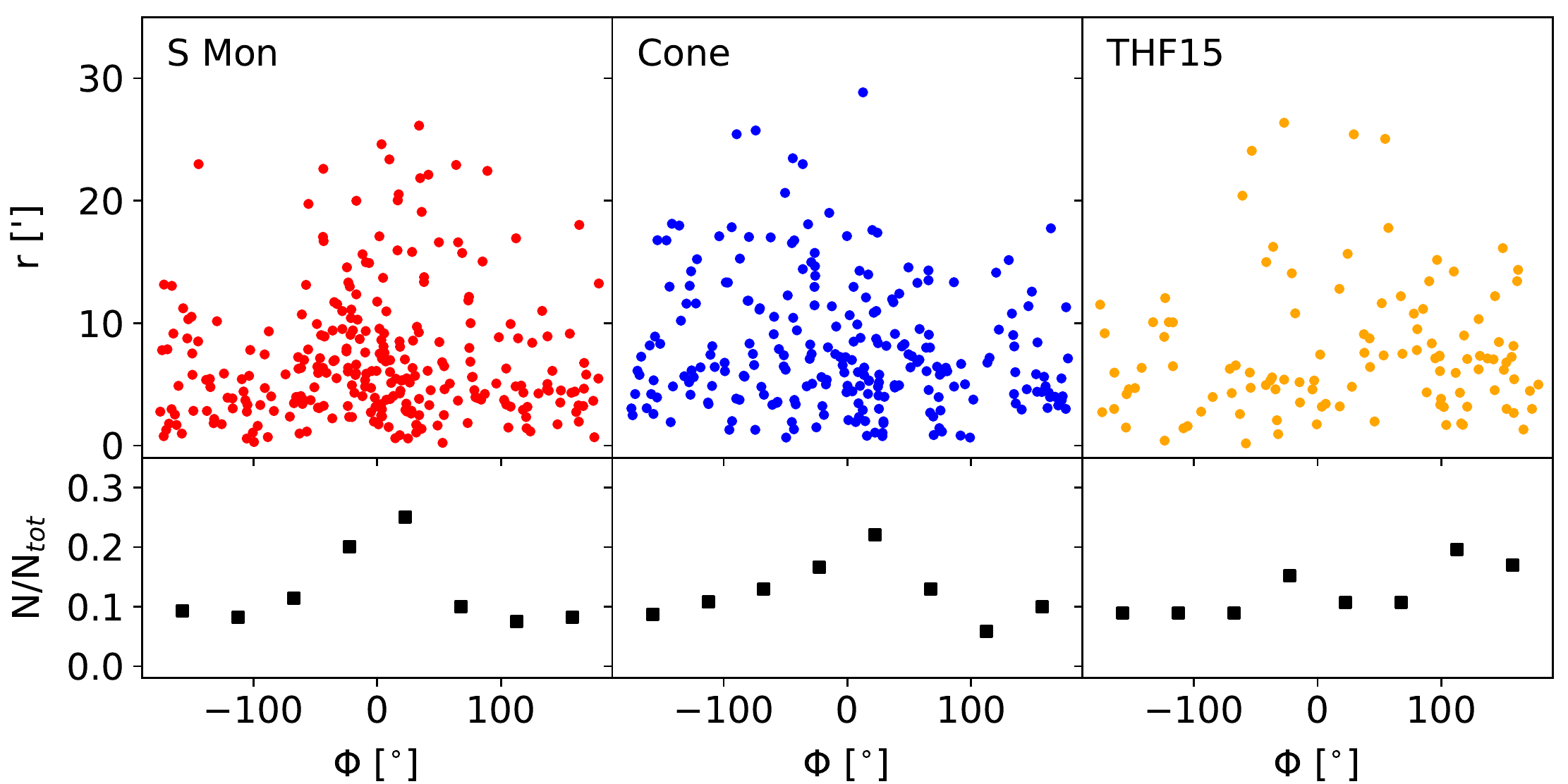}
\caption{Vectorial angle distributions of stars in the stellar groups of 
Mon OB1. The upper panels display the vectorial angles with respect 
to the projected radial distances from the central positions of host 
groups. The distributions of vectorial angles are shown by histograms 
in the lower panels. The histograms were obtained with a bin size 
of $45^{\circ}$. }\label{fig10}
\end{figure}
 
\subsection{Kinematics}\label{ssec:42}
\subsubsection{Internal kinematics in Mon OB1}\label{sssec:421}
We investigated the internal motions of stellar groups in Mon OB1. 
The upper left panel of Figure~\ref{fig9} yields the spatial 
distribution of stars along with the PM vectors (arrows) 
of three stellar groups relative to the systemic motion of Mon OB1, where 
the relative PM vectors are the median PM vectors of individual groups 
subtracted by the median PM of the entire system. The three groups 
do not have any significant motion along declination. The S Mon and 
THF15 groups are moving eastward with similar velocities, while 
the Cone group is moving westward at a larger velocity.

The relative PM vectors of individual members within a given 
group were obtained after subtracting their median PM. 
Figure~\ref{fig9} displays the relative PM vectors of members in 
the S Mon (upper right), Cone (lower left), and THF15 
(lower right) groups. Many members in the S Mon and Cone 
groups tend to show outward motions from the central position 
of each group, while there is no clear pattern of expansion 
in the THF15 group. 

In order to quantitatively probe the internal motions of stars, 
we measured the vectorial angles of members as used 
in our previous studies \citep{LNGR19,LHY20,LNH21}. The vectorial 
angle ($\Phi$) is an angle between the position vector from 
the group center and the relative PM of a given star. A  
zero value means that a star is radially escaping from its host group. 
We present the $\Phi$ distribution of individual members 
belonging to each group in Figure~\ref{fig10}. 

The $\Phi$ values of members in the S Mon group are 
clustered around 0$^{\circ}$, which is indicative of expansion. 
The Cone group also shows a pattern of expansion 
as many members of this group have $\Phi$ values around 
0$^{\circ}$. On the other hand, the members of THF15 
do not show clear outward motions given that there is 
no strong peak around $\Phi = 0^{\circ}$. 

Recent studies also detected expansion for the S Mon 
group \citep{KHS19,BKG20}, but not for the southern groups. 
The southern stellar groups are seen as a single cluster 
in optical passbands \citep{SBC08}, while two well-defined 
subgroups, the so-called Cone(C) and Spokes \citep{TLY06,SSB09}, 
are found around the embedded YSOs NGC 2264 IRS 1 and IRS 2 
in infrared passbands. The fact that a large fraction of 
YSOs in the Cone(C) and Spokes groups were not found 
in optical CMD indicates that these 
two groups are deeply embedded \citep{SB10}. The presence of 
molecular clouds toward the southern region can directly 
be confirmed from \citet{THF15}. Therefore, we speculate 
that there are four stellar groups (Cone, THF15, Cone(C), 
and Spokes) along the line of sight at the south of 
Mon OB1.

If this is true, \citet{KHS19} might have probed 
small parts of the Cone and THF15  groups, not actually 
the Cone(C) and Spokes groups. Then, the 
investigators could not find any pattern of expansion.

 \begin{figure}[t]
\epsscale{1.0}
\plotone{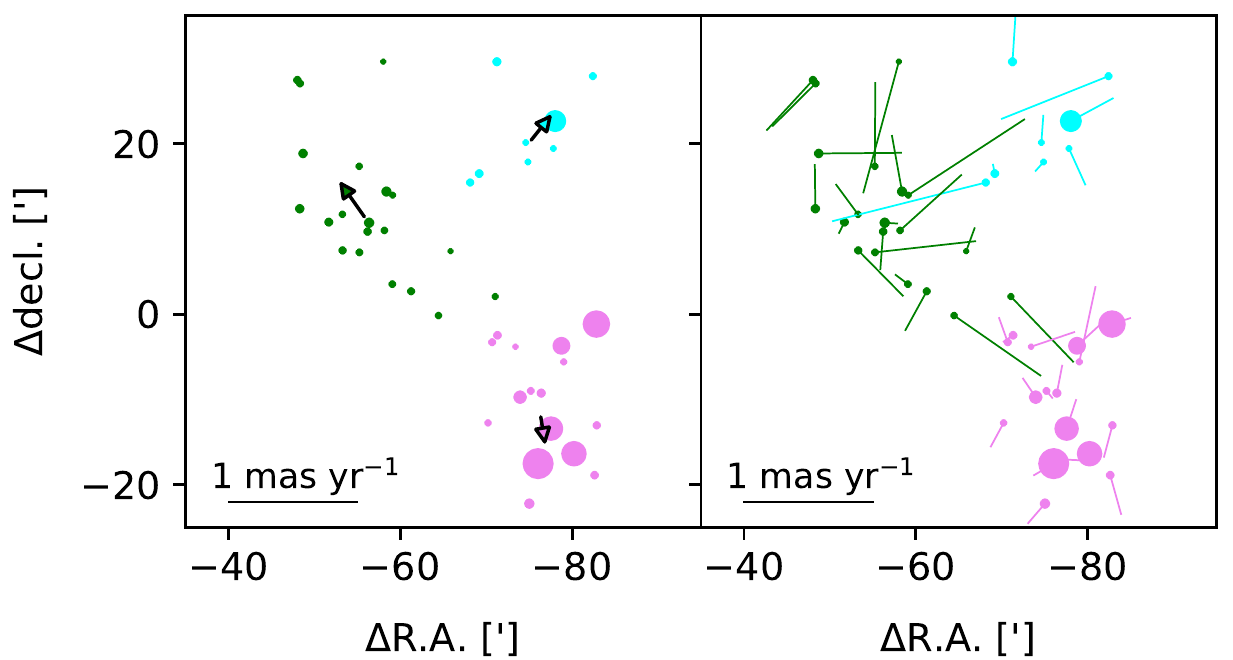}
\caption{Relative PMs of stars in Mon R1. The left panel 
shows the spatial distribution of three stellar groups. The systemic 
motions of these groups relative to the median PM of Mon R1 are 
shown by arrows. We plot the PM vectors (solid lines) of 
stars relative to the systemic motion of a given group in the right 
panel. The size of dots is scaled by the brightness of individual 
stars. The colors of the symbols are the same as those in 
Figure~\ref{fig7}.}\label{fig11}
\end{figure}
 
\subsubsection{Internal kinematics in Mon R1}\label{sssec:422}
We investigated the kinematics of stellar groups in Mon R1 in 
the same way. The left panel of Figure~\ref{fig11} displays the 
motions of the three groups relative to the systemic motion of 
Mon R1. IC 447 is moving south, while N2245/47 and IC 446 are 
moving northeast and northwest, respectively. These groups are 
receding away from each other. 

The PM vectors of individual members relative to their host 
groups are shown in the right panel. Although it is difficult 
to define the center of each group due to the small number of 
stars, members tend to show outward motions. Figure~\ref{fig12} 
shows the $\Phi$ distributions. Stars 
in the three groups have $\Phi$ values around 0$^{\circ}$, indicating 
that the members of these groups are scattered radially outward.

\begin{figure}[t]
\epsscale{1.0}
\plotone{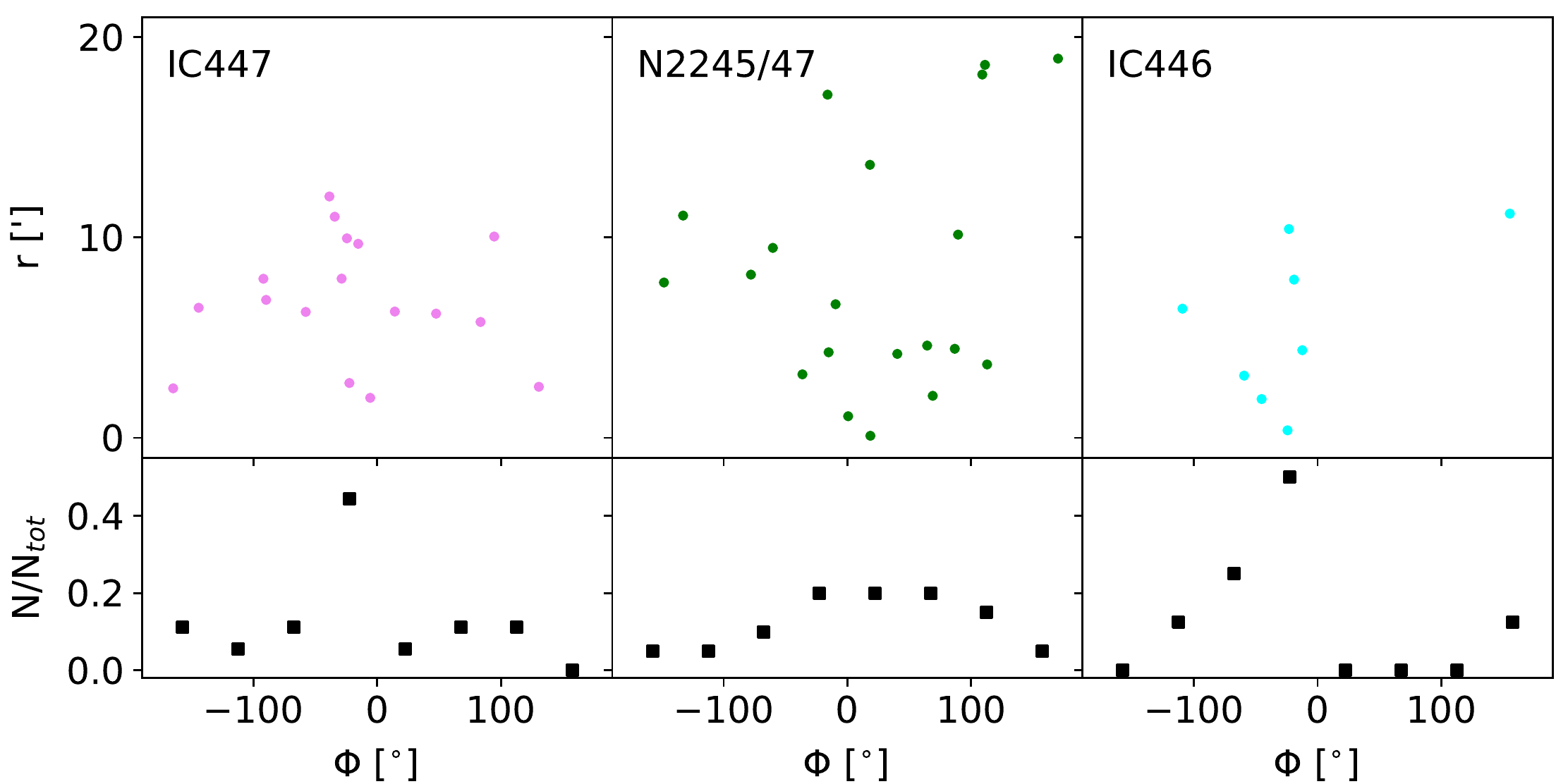}
\caption{Vectorial angle distributions of stars in the stellar groups of 
Mon R1. The upper panels display the vectorial angles with respect 
to the projected radial distances from the centers of IC 447, N2245/47, 
and IC 446 (from left to right), respectively. The distributions of 
vectorial angles are shown by histograms in the 
lower panels. The size of bins is $45^{\circ}$.  }\label{fig12}
\end{figure}

\subsection{Rotation}\label{ssec:43}
We searched for the signature of rotation for the three groups in 
Mon OB1 using the RVs of members as done by 
previous studies \citep{LKL09,MDFY13,LRN19,LNH21}. 
A projected rotational axis passing through the central position 
of a given group was set at a position angle of $0^{\circ}$ 
(from north to south) in the projected sky plane. 
We computed the difference between the mean RVs of stars in the two 
areas separated by the axis. The same computation was 
repeated for various position angles between 0$^{\circ}$ and 
360$^{\circ}$ with an interval of 20$^{\circ}$ in a counterclockwise 
direction (north to east). If a given cluster 
is rotating, the mean RV differences appear as a sinusoidal 
curve. 

\begin{figure}[t]
\epsscale{1.0}
\plotone{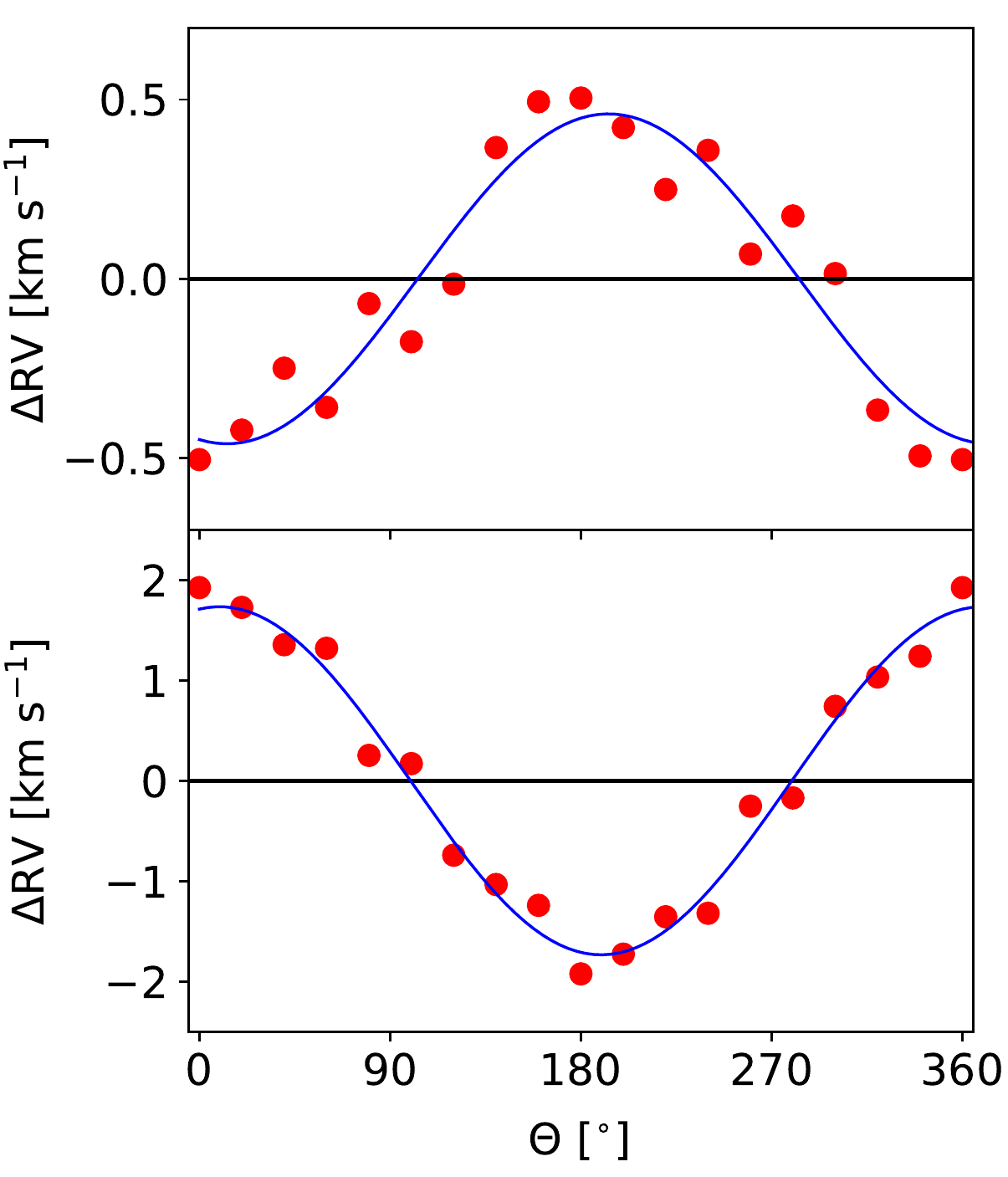}
\caption{Signature of rotation of the S Mon group (upper) and Cone group 
(lower). $\Delta$RV denotes the difference 
of mean RVs between two regions separated by a projected 
rotational axis at a given position angle ($\Theta$). The blue solid lines 
represent the best-fit sinusoidal curves. Half of the amplitude 
corresponds to the projected rotational velocity. }\label{fig13}
\end{figure}

We found the signature of rotation for the members within 
a projected radius of $7^{\prime}$ for the S Mon group and 
within $5^{\prime}$ for the Cone group. Figure~\ref{fig13} 
exhibits the variations of the mean RV differences with respect 
to position angles. These observed variations were fit by the 
sinusoidal curve : 

\begin{equation}
\Delta\langle\mathrm{RV}\rangle = 2 V_{\mathrm{rot}}\sin i \sin(\Theta + \Theta_0)
\end{equation}

\noindent where $V_{\mathrm{rot}}$, $i$, and $\Theta_0$ represent 
the rotational velocity, inclination angle of a rotational axis along 
the line of sight, and phase, respectively. The amplitude of the best-fit 
sinusoidal curve corresponds to twice the projected rotational velocity 
($V_{\mathrm{rot}} \sin i$). The S Mon and Cone groups are rotating 
at $0.23 \pm 0.02$ km s$^{-1}$ and $0.87 \pm 0.03$ km s$^{-1}$, 
respectively, if we assume an inclination angle of $90^{\circ}$. 
The position angle of the projected rotational axis can be 
estimated from $270^{\circ} - \Theta_0$.
The projected rotational axes of the S Mon and Cone groups almost lie 
from north to south, but these groups are rotating in opposite directions. 
On the other hand, we could not find any signature of rotation for the THF15 
group. The same method could not be applied to the stars in Mon R1 because 
of a small number of stars.

\begin{figure}[t]
\epsscale{1.0}
\plotone{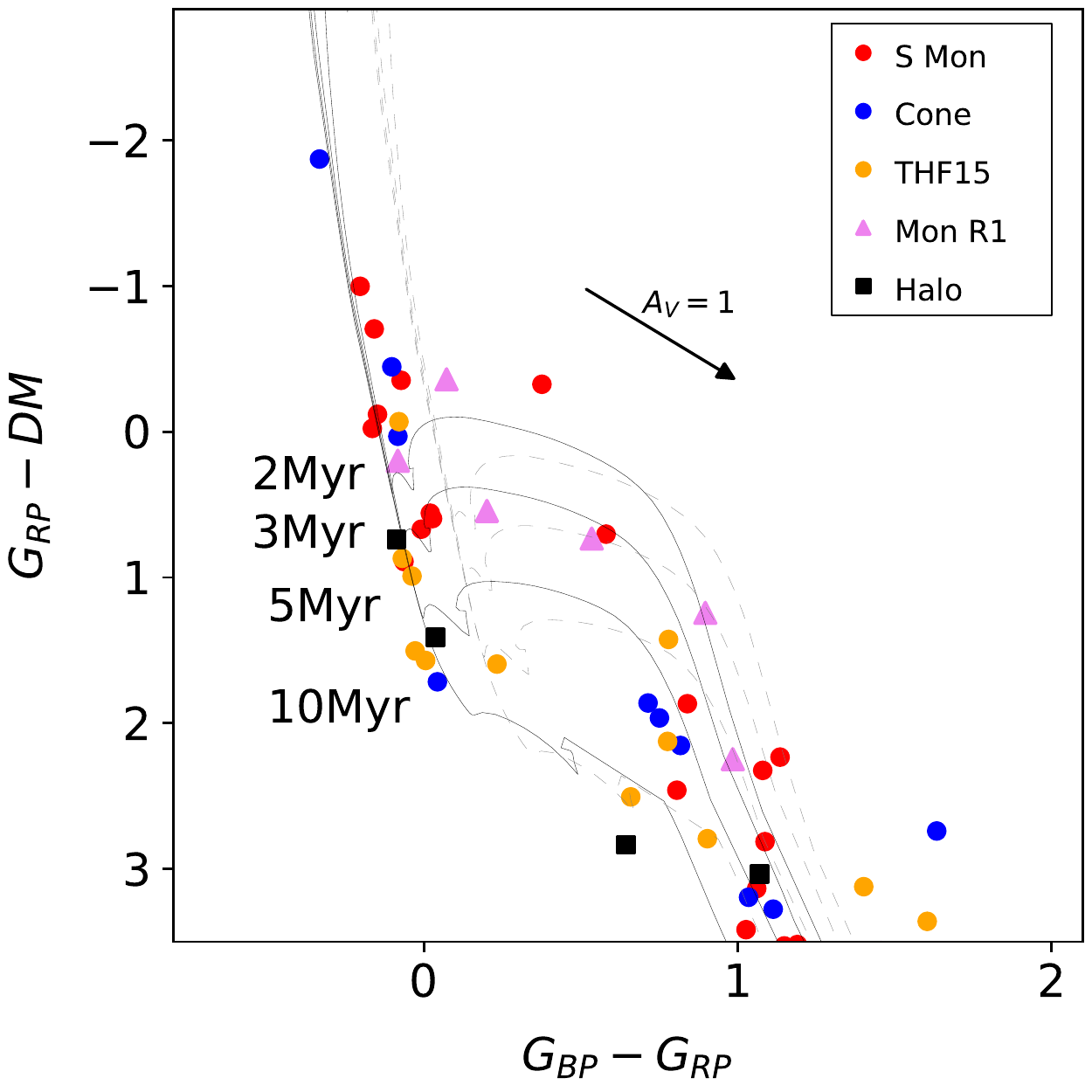}
\caption{Color-magnitude diagram of stars in the survey region. 
The meaning of the symbols is shown in the upper-right corner 
of the panel. The $G_{\mathrm{RP}}$ magnitude of individual stars 
were corrected by the distance moduli of the associated SFRs. The black 
and gray dashed curves show the isochrones reddened by 
$A_V$ of 0.22 and 0.62 mag, respectively, for ages of 2, 3, 5, and 
10 Myr \citep{D16,CDC16}. The black arrow represents the reddening 
vector corresponding to the total extinction of $A_V = 1.$ mag.}\label{fig14}
\end{figure}

\subsection{Ages of stellar groups}\label{ssec:44}
Figure~\ref{fig14} displays the CMD of stars in Mon OB1, Mon R1, and 
the halo. The $G_{\mathrm{RP}}$ magnitudes of members were 
corrected by the distance moduli of their host groups (see 
Section~\ref{sec:3}). We superimposed four isochrones (gray curves) 
for 2, 3, 5, and 10 Myr \citep{D16,CDC16} on the CMD. A mean total extinction ($A_V$)
of 0.22 mag in visual magnitude \citep{SBL97} was applied to the 
isochrones. The faint part of the CMD is significantly affected by 
some factors such as large photometric errors, high 
internal extinction of disk-bearing stars, and variabilities 
\citep[etc]{CSB14,SCB14,LSB15,SCR16}. In addition, 
the systematic difference between the isochrones from the adopted 
evolutionary models and the observed CMD is found for late-type 
pre-main sequence stars. For these reasons, we only considered 
the bright part of the CMD as seen in Figure~\ref{fig14}.

The magnitude (or luminosity) of the main sequence turn-on is 
sensitive to the age of a given coeval stellar system. The 
ages of members in Mon OB1 roughly range from 2  to 10 Myr. An 
age spread of 4 -- 5 Myr has been expected from the lithium 
abundances of pre-main sequence stars \citep{LSK16} and 
from the CMD analysis of \citet{VPS18}. Hence, star 
formation in Mon OB1 might be sustained on a several 
Myr scale. The S Mon group (2 -- 3 Myr) seems to be younger than the 
Cone and THF15 groups ($\gtrsim$ 5 Myr) given that main sequence 
turn-on appears at higher luminosity. The members of Mon R1 
have ages similar to those of the S Mon group. On the other hand, 
the CMD of the halo stars overlaps with that of the older populations 
of Mon OB1 (Cone and THF15). 

The photometric errors of stars around the main sequence 
turn-on ($G_{RP} < 12$ mag) are much smaller than 0.01 mag. 
Therefore, these are not the major source of uncertainties in age estimation. 
The error on distance ($\pm 40$ pc) corresponds to $\pm$0.1 
mag in distance modulus. The contribution of this error to age 
estimation is also negligible. The other factor is differential 
reddening across the survey region. 

\begin{figure*}[t]
\epsscale{1.0}
\plotone{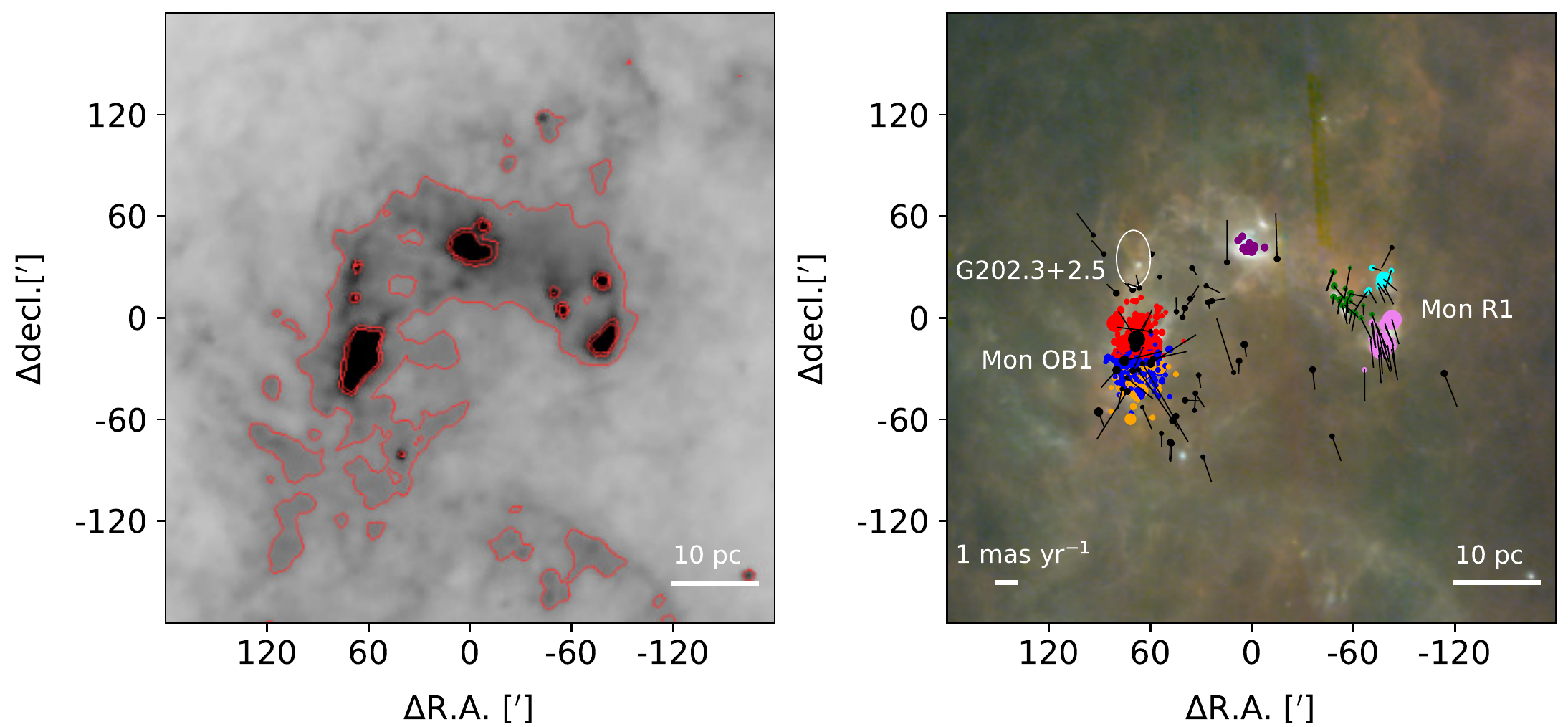}
\caption{Distributions of stars and gas in our survey region. 
The left and right panels display the IRIS image at 100$\micron$ \citep{ML05} 
and the AKARI false-color image (blue : 65 $\micron$, green : 90 $\micron$, 
and red : 140 $\micron$ -- \citealt{DTO15}), respectively. The contours 
in the left panel trace the arc-like structure of the remaining 
cloud. In the right panel, the PM vectors of stars relative to the  systemic motion of Mon OB1 
are shown by solid lines. The black dots represent the halo stars. 
The white ellipse represents the position of the SFR G202.3+2.5. 
Some infrared and submillimetric sources \citep{BNH88,DJK08} 
in the northern knot were marked by purple dots.  The colors of 
the other symbols are the same as in Figures~\ref{fig5} and \ref{fig7}. }\label{fig15}
\end{figure*}

The differential reddening over Mon OB1 is known to be 
small $\langle E(B-V) \rangle = 0.07 \pm 0.03$ \citep{SBL97}. 
\citet{SBC08} found higher reddening values of $E(B-V) \sim 
0.2$ for low-mass pre-main sequence stars in the region 
(see also \citealt{RMS02}). We plotted the isochrones 
reddened by this high reddening value (gray dashed curves) 
in Figure~\ref{fig14}. But, the colors and magnitudes of 
main sequence stars are closer to those of the 
isochrones reddened by the mean reddening value of 
$E(B-V) = 0.07$ ($A_V = 0.22$ mag). This implies that 
the differential reddening is not high enough to affect the 
relative ages among the stellar groups in Mon OB1.

Most of the members of Mon R1 seen in 
Figure~\ref{fig14} are the bright members of IC 447. The reddening 
of these stars due to the intracluster medium may be small. Indeed, 
they were found in the cavity of the dusty cloud (see figure 1 of 
\citealt{BDP20}). The B-type member with the bluest color in 
IC 447 has a color similar to those of members in Mon OB1. 
Therefore, the minimum reddening toward Mon R1 may be 
similar to the mean reddening of Mon OB1. There is a bright 
member of IC 446 in the CMD ($G_{RP} - DM = 1.24$, 
$G_{BP} - G_{RP} = 0.90$). The age of the star is about 2 Myr, 
which is comparable to the ages of the IC 447 members. The 
ages of the bright stars in IC 446 and IC 447 may not be 
significantly altered ($\lesssim$ 1 Myr) by differential reddening 
given the reddening vector.

\subsection{A Large-scale Distribution of Stars and Gas}\label{ssec:45}
We plotted the IRIS image at 100$\micron$ and the AKARI false-color 
(blue : 65 $\micron$, green : 90 $\micron$, and red :140 $\micron$) image 
of interstellar material over our survey region in Figure~\ref{fig15}. Interestingly, the 
IRIS image reveals a large arc-like structure across the survey region. 
The members of Mon OB1, Mon R1, and the halo were superimposed 
on the AKARI image. It seems clear that star formation is actively taking 
place along the arc-like structure. Indeed, bright knots notably host Mon OB and Mon R1, 
but also \object{G202.3+2.5} \citep{MJV19a,MJV19b} and some continuum 
sources that are bright at infrared and submillimeter wavelengths \citep{BNH88,DJK08}. 

Figure~\ref{fig15} also displays the PM vectors of members relative 
to the systemic motion of Mon OB1. A high fraction of the halo stars 
are found around Mon OB1 and tend to move outward from the 
association. Their $\Phi$ distribution has a strong peak 
at around 20$^{\circ}$. On the other hand, the members of Mon R1 
are systematically moving toward south relative to Mon OB1.

\section{Discussion}\label{sec:5}
\subsection{Implication on cluster formation}\label{ssec:51}
The S Mon and Cone groups show patterns of expansion 
as seen in the other stellar clusters \citep[etc]{CJW19,KHS19,LNGR19,
LHY20,LNH21}. About 50\% of members in the S Mon group are radially 
escaping from this group, but some members beyond $5^{\prime}$ 
($\Phi \sim \pm 180^{\circ}$, Figure~\ref{fig10}) are still sinking into 
the group center. The fraction of these members may be less 
than 20\% from the last two bins around $\Phi \sim \pm 180^{\circ}$ 
in the histogram of Figure~\ref{fig10}. The Cone group shows a similar pattern, but less clear 
than that of the S Mon group. Such a trend was also found in the 
Orion Nebula Cluster \citep{PRB20}. On the other hand, the 
young open clusters \object{IC 1805} and \object{NGC 2244} 
only show patterns of expansion without signature of collapse \citep{LHY20,LNH21}. 
The different internal kinematics among these clusters may result 
from the different initial conditions of cluster formation and evolution time. 

Many theoretical studies have tried to explain the expansion of stellar 
clusters as the result of their dynamical evolution after rapid gas expulsion \citep{T78,H80,LMD84,KAH01,BK13,BK15}. Our previous study \citep{LHY20} 
explained the expansion of the young open cluster \object{IC 1805} 
using an $N$-body simulation without the consideration of gas expulsion. 
This simulation considered a model cluster formed at an extremely subvirial 
state. The modelled cluster experienced collapse in the first 2 Myr and then 
expanded. As a result, the members of this clusters have $\Phi$ values 
around $\pm180^{\circ}$ within 2 Myr and then have $\Phi$ values 
around $0^{\circ}$ after the epoch of the major collapse.

The members of the S Mon and Cone groups have $\Phi$ 
distributions similar to the snapshots of the modelled cluster 
evolution during the transition epoch from collapse to rebound. 
Therefore, the monolithic cold collapse scenario can provide a 
possible explanation for the formation and evolution of these 
two groups. In addition, the rapid gas ejection and stellar 
feedback could affect the structure and dynamics of these 
groups \citep{GBR17}. 
 
The S Mon and Cone groups also show the signature of 
rotation. Some young stellar clusters were also found to be 
rotating, e.g., \object{R136} \citep{HGE12}, \object{Trumper 15} 
\citep{KHS19}, and \object{NGC 2244} \citep{LNH21}. These 
results provide important constraints on the cluster formation process, 
such as the monolithic collapse of rotating clouds or the 
hierarchical assembly of subclusters \citep{CLG17,M17}. 
A large fraction of molecular clouds in external galaxies were found 
to be rotating \citep{REPB03,T11,BHR20}. Rotating clusters 
can naturally form in such rotating clouds. However, the groups in Mon OB1 
have different directions of rotation, which cannot be explained 
by only the monolithic collapse of a single molecular cloud. 
 
Theoretically, collisions between molecular clouds can result in 
a larger cloud with retrograde rotation with respect to the galactic 
rotation on a large spatial scale \citep{DBP11}. If such collisions 
could occur between gaseous and stellar clumps on small spatial 
scales (several pc), then the angular momentum vector of the natal 
cloud could be changed \citep{M17}. In addition, the observational 
evidence for an on-going merger of stellar clusters was reported \citep{SLG12}. 
Therefore, our findings suggest that at least one of the groups 
in Mon OB1 might have been formed via the hierarchical merging process.

On the other hand, the stellar groups in Mon R1 do not seem 
to form as self-gravitating systems given the weak central 
concentration of stars. As these groups expand and move away 
from each other, their members will eventually disperse to become 
field star population in the Galactic disk \citep{MS78,BPS07}.

\begin{figure*}[t]
\epsscale{1.0}
\plotone{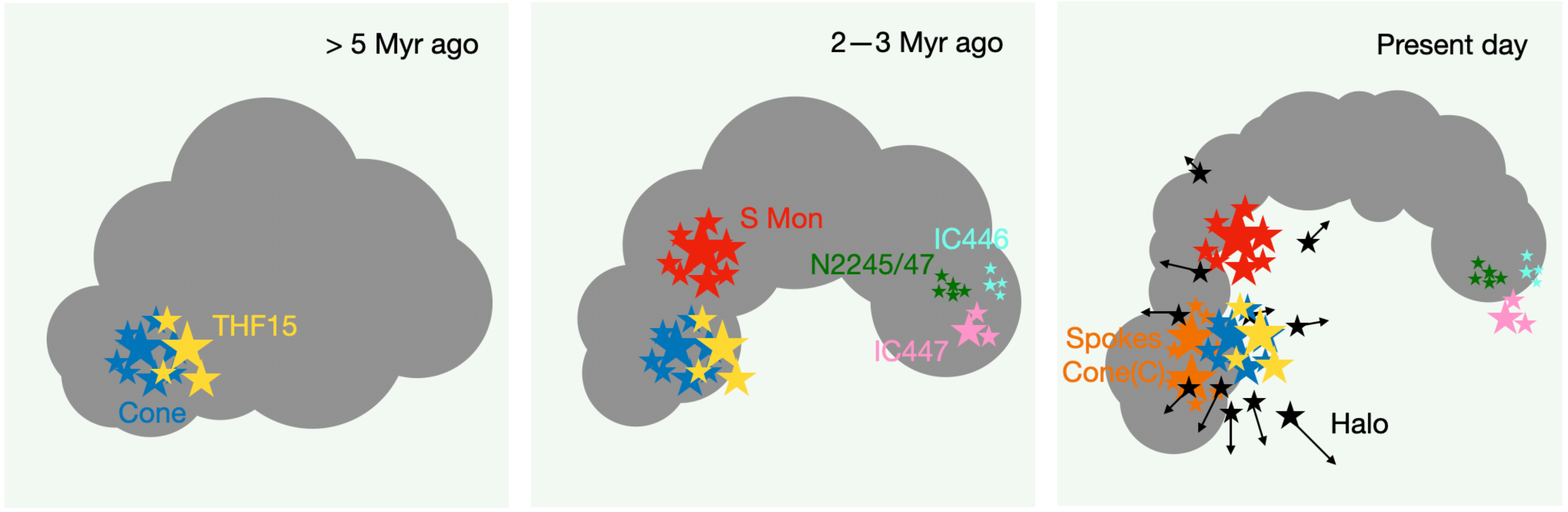}
\caption{Schematic sketch of star formation history in Mon OB1 
and Mon R1 from $>$ 5Myr ago (left) to present (right). Stellar 
groups are labelled at the epoch when they were formed. See 
the text for details.}\label{fig16}
\end{figure*}

\subsection{Star formation on different spatial scales}\label{ssec:52}
The projected distance and the line of sight distance between Mon 
OB1 and Mon R1 are about 20 and 40 pc, respectively, which is 
the typical size of giant molecular clouds. Members in 
these two associations have velocities in almost the same range 
(see Figures~\ref{fig6} and ~\ref{fig8}). As a comparison point, 
note that the Orion A cloud shows an RV variation larger than 
10 km s$^{-1}$ from north to south \citep{YLC21}. In addition, 
molecular gas constituting the arc-like structure as seen in 
Figure~\ref{fig15} was found in the same velocity channel 
($0 < V [km/s] < 15$ from figure 2-b of \citealt{OMT96}). 
Hence, the two associations and 
the other small SFRs related to bright knots have 
formed in the same molecular cloud. However, it is unclear how 
their formation is related to each other. 

Based on the age distributions of pre-main sequence stars in Mon OB1, 
\citet{SB10} proposed an outside-in star formation history for Mon 
OB1. Star formation has been initiated from the halo and propagated 
via the S Mon and Cone groups to the Cone(C) and Spokes 
groups around the embedded YSOs NGC 2264 IRS 1 and IRS 2 (see also 
\citealt{VPS18}). However, this scenario should be slightly 
modified, according to our results. Indeed, star formation initiated in 
the southern region (Cone and THF15) and then occurred in 
the northern region (S Mon). Recent star-forming activity has 
ignited in the molecular gas behind the southern region, 
which may have resulted in the formation of the Cone(C) and 
Spokes groups. The star formation history of Mon OB1 and 
Mon R1 is illustrated in a simple cartoon (Figure~\ref{fig16}).

A number of halo stars were found around Mon OB1, and they 
seem to be escaping from the association (Figure~\ref{fig15}). 
Most of them might thus have formed in Mon OB1 as the first 
generation of stars. The fact that the halo stars have almost the 
same ages as those of the older populations in Mon OB1 
supports this argument. Therefore, the eastern halo might 
have been formed by stars escaping from the association 
rather than the star formation {\it in-situ} proposed by \citet{SB10}.

The systemic motion of Mon R1 may provide clues to 
the formation of this association. If this association 
had been formed in the compressed clouds by feedback from the 
O-type binary S Monocerotis which is located to the east of Mon R1, 
its members would move westward. However, they do not show 
such a systematic motion toward west (Figure~\ref{fig15}). 
Furthermore, there is no significant 
age difference between the S Mon group and Mon R1. The stellar 
groups in Mon R1 might thus have spontaneously formed at about 
the same epoch as the formation of the S Mon group in Mon OB1.  

Herschel images at submillimeter wavelengths reveal the 
networks of filaments in many molecular clouds \citep{A15}. 
The presence of filamentary structures in molecular clouds 
has been accepted as a ubiquitous feature. Turbulence could 
play a key role in the formation of filamentary structures 
\citep{L81,PJG01}. Recently, there is increasing evidence 
that magnetic fields significantly contribute to the formation 
of filaments on a small scales \citep[etc]{WLE19,DHF20}. 
Cores and protostars form after the gravitational fragmentation 
of filaments \citep{AMB10}. Filament hubs with high densities 
are the sites of stellar cluster formation \citep{SCH12,GLZ13,
TFS19}. The relics of such filaments have been found 
toward the Vela OB2 association and the Orion region 
\citep{JBB19,BBJ20,PYT21}. In our survey region, 
SFRs including Mon OB1 and Mon R1 are 
distributed along a large arc-like structure in a hierarchical way. 
The formation of structures on different spatial scales 
by the actions of turbulence, gravity, and magnetic field 
may be the essential formation process of Mon OB1 and Mon R1. 

\section{Summary}\label{sec:6} 
In this study, we investigated the spatial distributions and 
kinematic properties of young stellar population in the two 
stellar associations Mon OB1 and Mon R1 to understand star 
formation process and their physical association. 

We first isolated member candidates in a 
$6^{\circ} \times 6^{\circ}$ survey region using the 
published data sets. Then, a total of 728 members were 
finally selected from the criteria based on the Gaia parallaxes 
and PMs. The spatial distributions of these stars show 
substructures that are kinematically distinct. Mon OB1 
contains three optically visible stellar groups, the 
S Mon, Cone, and THF15 groups. We also 
suggested the possibility that there are two embedded 
groups (Spokes and Cone(C)) behind the optically visible 
groups. Mon R1 hosts the open cluster IC 447 and two partially 
embedded groups (N2245/47 and IC 446). In addition, 
some stars were found in the halo region.

The stellar groups, except for \object{THF15} show patterns 
of expansion as seen in many associations. In addition, 
the signature of rotation was detected for the S Mon and 
Cone groups. Interestingly, these groups are rotating 
in opposite directions, which could be a trace if clouds 
having merged in the past.

We analyzed the CMD of members to infer the star 
formation history in the survey region. The members 
of Mon OB1 have ages ranging 2 Myr to $\gtrsim$ 5 Myr. 
The ages of the Mon R1 members (2 -- 3 Myr) are similar 
to those of the younger population in Mon OB1, 
while the halo stars ($\gtrsim$ 5 Myr) have ages similar 
to those of the older population. Furthermore, the motions in
Mon R1 are not pointing away from Mon OB1. This suggests 
that Mon OB1 and Mon R1 might have formed independently 
in a giant molecular cloud.

In addition, Mon OB1 and Mon R1 belong to a 
large scale arc-like structure comparable to the size 
of typical giant molecular clouds. Actually, more 
star formation activities on small scales are 
found along this large structure, forming a 
hierarchy: isolated stars, clusters, and 
associations. Hence, these two associations 
might have formed within the same cloud in a 
hierarchical way. In addition, the expansion 
of stellar groups plays a crucial role in the 
formation of the halo population.

\begin{acknowledgments}
This paper has made use of data obtained under the K-GMT Science Program 
(PID: GEMINI-KR-2020A-003 and Gemini program number: GS-2020A-Q-239) 
funded through Korean GMT Project operated 
by Korea Astronomy and Space Science Institute (KASI) and from the European 
Space Agency (ESA) mission {\it Gaia} (https://www.cosmos.esa.int/gaia), 
processed by the {\it Gaia} Data Processing and Analysis Consortium 
(DPAC, https://www.cosmos.esa.int/web/gaia/dpac/consortium). Funding 
for the DPAC has been provided by national institutions, in particular the institutions 
participating in the {\it Gaia} Multilateral Agreement. This research has also 
made use of the SIMBAD database,operated at CDS, Strasbourg, France. 
Based on observations obtained at the international Gemini Observatory, 
a program of NSF’s NOIRLab [ include additional acknowledgment here, 
see section below ], which is managed by the Association of Universities 
for Research in Astronomy (AURA) under a cooperative agreement with 
the National Science Foundation on behalf of the Gemini Observatory 
partnership: the National Science Foundation (United States), National 
Research Council (Canada), Agencia Nacional de Investigación y Desarrollo 
(Chile), Ministerio de Ciencia, Tecnolog\'ia e Innovaci\'on (Argentina), 
Minist\'erio da Ci\^encia, Tecnologia, Inova\c c\~oes e Comunicações (Brazil), 
and Korea Astronomy and Space Science Institute (KASI) (Republic of Korea).
This work used the Immersion Grating Infrared Spectrometer (IGRINS) 
that was developed under a collaboration between the University of Texas 
at Austin and the KASI with the financial support of the US National Science 
Foundation under grants AST-1229522 and AST-1702267, of the University 
of Texas at Austin, and of the Korean GMT Project of KASI. This work was 
supported by the National Research Foundation of Korea (NRF) 
grant funded by the Korean government (MSIT) (Grant No : 
NRF-2019R1C1C1005224 and 2022R1C1C2004102). Y.N. acknowledges support from the Fonds 
National de la Recherche Scientifique (Belgium), the European Space Agency 
(ESA) and the Belgian Federal Science Policy Office (BELSPO) in the 
framework of the PRODEX Programme (contracts linked to XMM and Gaia). 
\end{acknowledgments}

\vspace{5mm}
\facilities{Gemini-South:8.2m}

%% Similar to \facility{}, there is the optional \software command to allow 
%% authors a place to specify which programs were used during the creation of 
%% the manuscript. Authors should list each code and include either a
%% citation or url to the code inside ()s when available.

\software{{\tt xcsao} \citep{KM98}, {\tt SPECTRUM} \citep{GC94}, 
{\tt IGRINS pipeline 2} \citep{LGK17}, {\tt NumPy} \citep{HMvdW20}, {\tt Scipy} \citep{VGO20}}

%% Appendix material should be preceded with a single \appendix command.
%% There should be a \section command for each appendix. Mark appendix
%% subsections with the same markup you use in the main body of the paper.

%% Each Appendix (indicated with \section) will be lettered A, B, C, etc.
%% The equation counter will reset when it encounters the \appendix
%% command and will number appendix equations (A1), (A2), etc. The
%% Figure and Table counter will not reset.

%\appendix

%\section{Appendix information}

%\bibliography{sample631}{}
%\bibliographystyle{aasjournal}

%% This command is needed to show the entire author+affiliation list when
%% the collaboration and author truncation commands are used.  It has to
%% go at the end of the manuscript.
%\allauthors

%% Include this line if you are using the \added, \replaced, \deleted
%% commands to see a summary list of all changes at the end of the article.
%\listofchanges

\end{document}